%% file: main.tex
\documentclass[aps,prd,twocolumn,superscriptaddress,amsmath,amssymb]{revtex4}

\usepackage{graphicx} % Include figure files
\usepackage{epstopdf} % Allow to include eps files
\usepackage{dcolumn} % Align table columns on decimal point
\usepackage{xcolor} % Color support
\usepackage{amssymb} % Extended symbol collection
\usepackage{hyperref} % Generate pdfs with hyperlinks
\usepackage[utf8]{inputenc} % Allow UTF8 characters
\usepackage[T1]{fontenc}
\usepackage[english]{babel} % All publications are in English
\usepackage[displaymath, mathlines]{lineno} % Line numbers for drafts
\usepackage{blindtext} % For testing
\usepackage{amsmath}
\usepackage{siunitx}
\usepackage{orcidlink}
\usepackage{colordvi}
\usepackage{color}
\usepackage{amssymb}
\usepackage{url}
\graphicspath{{figures}}
\usepackage{tabularx}
\usepackage{comment}
\usepackage{multirow}

\input{definitions}

\begin{document}

{\onecolumngrid
\noindent
KEK preprint: 2023-55 \\
Belle II preprint: 2024-007\\
}

\title{Search for a $\mu^+\mu^-$ resonance in four-muon final states at Belle II }

  \author{I.~Adachi\,\orcidlink{0000-0003-2287-0173}} % 2590
  \author{K.~Adamczyk\,\orcidlink{0000-0001-6208-0876}} % 2239
  \author{L.~Aggarwal\,\orcidlink{0000-0002-0909-7537}} % 10083
  \author{H.~Ahmed\,\orcidlink{0000-0003-3976-7498}} % 11323
  \author{H.~Aihara\,\orcidlink{0000-0002-1907-5964}} % 2223
  \author{N.~Akopov\,\orcidlink{0000-0002-4425-2096}} % 9443
  \author{A.~Aloisio\,\orcidlink{0000-0002-3883-6693}} % 2194
  \author{N.~Anh~Ky\,\orcidlink{0000-0003-0471-197X}} % 2218
  \author{D.~M.~Asner\,\orcidlink{0000-0002-1586-5790}} % 4684
  \author{H.~Atmacan\,\orcidlink{0000-0003-2435-501X}} % 2538
% ...
% 2387
\collaboration{The Belle II Collaboration}
\noaffiliation

\begin{abstract}

We report on a search for a resonance $X$ decaying to a pair of muons in  $e^{+}e^{-}\rightarrow \mu^+ \mu^- X$ events in the 0.212--9.000\gevcc\ mass range, using  178\invfb of data  collected by the Belle~II experiment at the SuperKEKB collider at a center of mass energy of 10.58\gev.
The analysis probes two different models of $X$ beyond the standard model: a \zprime\ vector boson in the \lmultau\ model and a muonphilic scalar.
We observe no evidence for a signal and set exclusion limits at the 90\% confidence level on the products of cross section and branching fraction for these processes, ranging from 0.046\fb to 0.97\fb for the \lmultau\ model and from 0.055\fb to 1.3\fb for the muonphilic scalar model. For masses below 6\gevcc, the corresponding constraints on the couplings of these processes to the standard model range from 0.0008 to 0.039 for the \lmultau\ model and from 0.0018 to 0.040 for the muonphilic scalar model. These are the first constraints on the muonphilic scalar from a dedicated search. 

\end{abstract}

\maketitle

\section{Introduction}\label{sec:introduction}
The standard model (SM) of particle physics is a highly predictive theoretical framework describing fundamental particles and their interactions. Despite its success, the SM is known to provide an incomplete description of nature.
For example,  it does not address the phenomenology related to dark matter, such as the observed relic density~\cite{BERTONE2005279}.  
In addition, some experimental observations show inconsistencies with the SM. 
Prominent examples include the longstanding difference between the measured and the expected value of the muon anomalous magnetic-moment $(g-2)_{\mu}$~\cite{PhysRevD.73.072003,  AOYAMA20201, aguillard2023measurement}, possibly reduced by expectations based on lattice calculations \cite{BMWLattice}, and the tensions in flavor observables reported by the \textit{BABAR}, Belle, and LHCb experiments~\cite{BaBar:2013mob, LHCb:2017rln, Belle:2019rba}. 
Some of these observations can be explained with the introduction of additional interactions, possibly lepton-universality-violating, mediated by non-SM neutral bosons~\cite{SALA2017205, Chen:2017usq, Greljo:2021xmg}.  
Examples include the  \lmultau\ extension of the SM and a muonphilic scalar model. 

The \lmultau\ extension of the SM~\cite{PhysRevD.43.R22,Shuve:2014doa, Altmannshofer:2016jzy} gauges the difference between the muon and the $\tau$-lepton  numbers, giving rise to a new massive, neutral vector boson, the \zprime. Among the SM particles, this particle couples only to $\mu$, $\tau$, $\nu_\mu$, and $\nu_\tau$, with a coupling constant $g^{\prime}$. The \zprime\ could also mediate interactions between SM and dark matter.

The muonphilic scalar $S$ is primarily proposed as a solution for the  $(g-2)_{\mu}$ anomaly~\cite{harris2022snowmass, gori2022dark, forbes2023new, Capdevilla_2022}. This particle couples exclusively to muons through a Yukawa-like interaction, which 
is not gauge-invariant under the SM gauge symmetry and may arise from a high-dimension operator term at a mass scale beyond the SM.  In contrast to the \lmultau\ model, the muonphilic scalar model needs a high-energy completion.

Searches for a $Z^{\prime}$ decaying to muons have been reported by the \textit{BABAR}~\cite{TheBABAR:2016rlg}, Belle~\cite{PhysRevD.106.012003}, and CMS~\cite{cms} Collaborations. An invisibly decaying \zprime\ has been searched for by the Belle~II~\cite{PhysRevLett.124.141801, zpinvis_update}
and NA64-$e$~\cite{PhysRevD.106.032015} experiments.
The Belle~II experiment also searched recently for a \zprime\ decaying to $\tau^+\tau^-$~\cite{xtotautau}.
Constraints on the existence of a muonphilic scalar have been obtained by reinterpretations of \zprime\ searches into muons~\cite{Capdevilla_2022}. However, important experimental details may be unaccounted for in these reinterpretation studies, including the significantly different kinematic properties of the signal and the corresponding variation of the efficiency. 

\begin{figure}
\includegraphics[width=0.98\linewidth]{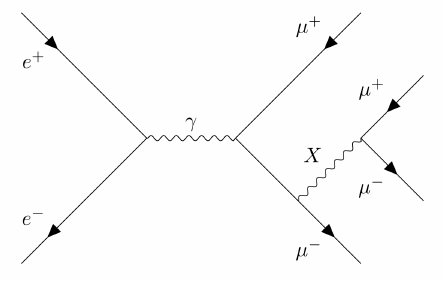}
\caption{\label{fig:Zp_diagram}Leading-order Feynman diagram for the process $e^+e^-\to\mu^+\mu^-X, X\to\mu^+\mu^-$.}
\end{figure}

Here we report a search for the process $e^+e^- \to \mu^+\mu^- X$, with $X \to \mu^+ \mu^-$, where $X$ indicates $Z^\prime$ or $S$. The signal signature is a narrow enhancement in the mass distribution of oppositely charged muons
\Mmumu\ in $e^+e^- \to \mu^+ \mu^- \mu^+ \mu^-$ events. 
We use data collected by the Belle II experiment at a center-of-mass (c.m.) energy $\sqrt{s}$ corresponding to the mass of the $\Upsilon(4S)$ resonance. 
The \lmultau\ model is used as a benchmark to develop the analysis; we then 
apply the same selections to the muonphilic scalar model and evaluate the performance.
In both models, the $X$ particle is at leading order emitted as final-state radiation (FSR) from one of the muons, as shown in Fig.~\ref{fig:Zp_diagram}.
For the range of couplings explored in this study, the lifetime of $X$ is negligible compared to the experimental resolution.
The analysis techniques are optimized using simulated events prior to examining data. 

\begin{figure}
\includegraphics[width=0.98\linewidth]{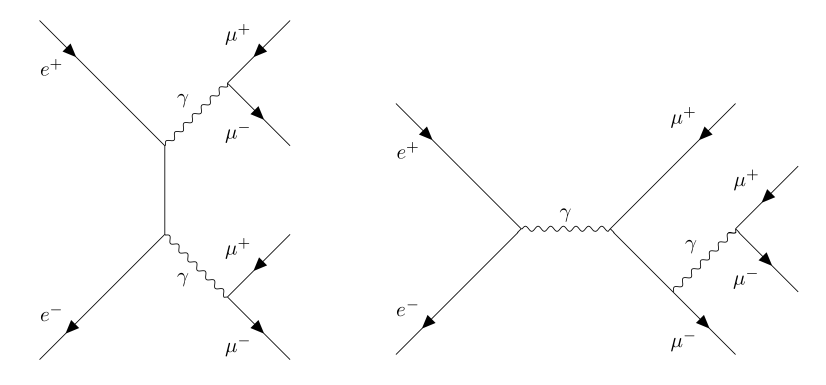}
\caption{\label{fig:bkg_diag} Leading-order Feynman diagrams for the two main contributions to the \fourmu\ SM background: double photon conversion (left) and annihilation (right).}
\end{figure}

We select events with exactly four charged particles with zero net charge, where at least three are identified as muons, with an invariant mass \mfourmu\ close to $\sqrt{s}/c^2$, and with negligible detected energy in addition to that associated to the charged particles. The dominant, non-peaking background is the SM \fourmu\ process, whose main production diagrams are shown in Fig.~\ref{fig:bkg_diag}. The analysis uses kinematic variables combined with a multivariate technique  to enhance the signal-to-background ratio.  
A kinematic fit improves the dimuon mass resolution. 
The signal yield is extracted through a series of fits  to the \Mmumu\ distribution, which allows an estimate of the background directly from data. 

The paper is organized as follows. In Sec.~\ref{sec:experiment} we briefly describe the Belle II experiment. In Sec.~\ref{sec:data} we report the datasets and the simulation used. In Sec.~\ref{sec:selections} we present the  event selections. In Sec.~\ref{sec:fit} we describe the signal modeling and the fit technique to extract the signal. In Sec.~\ref{sec:syst} we discuss the systematic uncertainties. In Sec.~\ref{sec:results} we describe and discuss the results. Section~\ref{sec:summary} summarizes our conclusions.

\section{The Belle II experiment}\label{sec:experiment}
The Belle~II detector~\cite{Abe:2010sj, ref:b2tip}
consists of several subdetectors arranged in a cylindrical structure around the $e^{+}e^{-}$ interaction point. 
The longitudinal direction, the transverse plane, and the polar angle $\theta$ are defined with respect to the detector's cylindrical axis in the direction of the electron beam.

Subdetectors relevant for this analysis are briefly described here in order from innermost out; a full description of the detector is given in Refs.~\cite{Abe:2010sj, ref:b2tip}. The innermost subdetector is the vertex detector, which consists of two inner layers of silicon pixels and four outer layers of silicon strips. The second pixel layer was only partially installed for the data sample we analyze, covering one sixth of the azimuthal angle. 
The main tracking subdetector is a large helium-based small-cell drift chamber. The relative charged-particle transverse momentum resolution, $\frac{\Delta p_T}{p_T}$, is typically 0.1\%$p_T$ $\oplus$ 0.3\%, with $p_T$ expressed in GeV/$c$. 
Outside of the drift chamber, time-of-propagation and aerogel ring-imaging Cherenkov detectors provide charged-particle identification in the barrel and forward end cap region, respectively. 
An electromagnetic calorimeter consists of a barrel and two end caps made of CsI(Tl) crystals: it reconstructs photons and identifies electrons. A superconducting solenoid, situated outside of the calorimeter, provides a 1.5~T magnetic field. A $K^0_L$ and muon subdetector (KLM) is made of iron plates, which serve as a magnetic flux-return yoke, alternated with resistive-plate chambers and plastic scintillators in the barrel and with plastic scintillators only in the end caps. 
In the following, quantities are defined in the laboratory frame unless specified otherwise.

\section{Data and simulation}\label{sec:data}
We use a sample of $e^{+}e^{-}$ collisions produced at c.m.\ energy $\sqrt{s}=10.58\gev$ in 2020--2021 by the SuperKEKB asymmetric-energy collider~\cite{superkekb} at KEK. The data, recorded by the Belle~II detector, correspond to an integrated luminosity of 178\invfb~\cite{lumi}.

Simulated signal \zmumu\ and \Smumu\ events are generated using
\texttt{MadGraph5\textunderscore aMC@NLO}~\cite{Alwall2014} with initial-state radiation (ISR) included~\cite{isr-plugin}.
Two independent sets of $Z^{\prime}$ events are produced, with \zprime\ masses, \zprimemass, ranging from 0.212\gevcc to 10\gevcc in steps of 250\mevcc, to estimate efficiencies, define selection requirements, and develop the fit strategy, and in steps of 5\mevcc, exclusively dedicated to the training of the multivariate analysis. 
Samples of $S$ events are generated in 40\mevcc steps for \Smass\ masses between 0.212\gevcc and 1\gevcc and in 250\mevcc steps from 1\gevcc to 10\gevcc.

Background processes are simulated using the following generators:
\fourmu, \eemumu, \foure, \mmtt\ and \eett, with \texttt{AAFH}~\cite{ref:fourlepton};
\mumugamma\  with \texttt{KKMC}~\cite{ref:kkmc};
\taugamma\ with \texttt{KKMC} interfaced with \texttt{TAUOLA}~\cite{ref:tauola};
\eepipi\ with \texttt{TREPS}~\cite{uehara2013treps};
\pipigamma\ with \texttt{PHOKHARA}~\cite{ref:phokhara};
\eegamma\ with \texttt{BabaYaga@NLO}~\cite{ref:babayaga};
$e^{+}e^{-} \to u \bar u, d \bar d, s \bar s, c \bar c$ with \texttt{KKMC} interfaced with \texttt{\textsc{Pythia8}}~\cite{pythia8} and \texttt{\textsc{EvtGen}}~\cite{evtgen} and $e^{+}e^{-} \to B^0\bar{B}^0$\ and $e^{+}e^{-}\to B^+B^-$\ with \texttt{\textsc{EvtGen}} interfaced with \texttt{\textsc{Pythia8}}.
Electromagnetic FSR is simulated with {\texttt{\textsc{Photos}}}~\cite{Barberio:1990ms, Barberio:1993qi} for processes generated with {\tt \textsc{EvtGen}}.
The \texttt{AAFH} generator, used for the four-lepton processes, including the dominant \fourmu\ background, does not simulate ISR effects.
This is a source of disagreement between data and simulation.
Other sources of nonsimulated backgrounds include \mumupp\ and more generally $e^{+}e^{-}\rightarrow  \mu^+\mu^- h$ and $e^{+}e^{-}\rightarrow  \pi^+\pi^- h$, where $h$ is typically a low-mass hadronic system; 
$e^{+}e^{-}\rightarrow  J/\psi \; \pi^+\pi^-$ with $J/\psi \to \mu^+ \mu^-$; $e^{+}e^{-}\rightarrow  \psi(2S) \gamma$ with $\psi(2S) \to J/\psi \; \pi^+\pi^-$ and $J/\psi \to \mu^+ \mu^-$; and $e^{+}e^{-}\rightarrow  \Upsilon(nS) \; \pi^+\pi^-$ with $\Upsilon(nS) \to  \mu^+ \mu^-$ and $n=1,2,3$. 

The detector geometry and interactions of final-state particles with detector material are simulated using \texttt{\textsc{Geant4}}~\cite{ref:geant4} and the Belle~II software~\cite{basf2, basf2-zenodo}.

\section{Selections}\label{sec:selections}
The selection requirements are divided into four categories: trigger, particle identification, candidate selections, and final background suppression.  
\subsection{Trigger selections}\label{sec:trigger}
We filter events selected by
the logical OR of a three-track trigger and a single-muon trigger. 
The efficiency of both triggers is measured using a reference calorimeter-only trigger, which requires a total energy deposit above 1\gev in the polar angle region $22^\circ < \theta < 128^\circ$. 
We require a single electron of sufficient energy to activate the calorimeter trigger. 
The three-track trigger requires the presence of at least three tracks with $37^\circ < \theta < 120^\circ$. The efficiency of this trigger is measured in four-track events containing at least two pions and one electron and depends on the transverse momenta $p_T$ of the two charged particles with lowest transverse momenta, reaching a plateau close to 100\% for $p_T$ above 0.5\gevc.  
The single-muon trigger is based on the association of hits in the barrel KLM with geometrically matched tracks extrapolated from the inner tracker. The efficiency of this trigger is measured in a sample of two-track events with one electron and one muon, mostly from the \tautau\ process, reaching a plateau of about 90\% in the polar angle range $51^\circ < \theta < 117^\circ$.
The efficiency for events with multiple muons is computed using the single-muon efficiency assuming no correlation. 
The overall trigger efficiency is 91\% for \zprimemass\ close to the dimuon mass, increases smoothly to a plateau close to 99\% in the mass range 2.5--8.5\gevcc, and then drops to 89\% at 10\gevcc.  It is slightly higher, 95\%,  for low masses in the $S$ case, due to the harder spectrum of the muonphilic scalar (see Sec.~\ref{sec:efficiency}). 

\subsection{Particle identification}\label{sec:muid}
The identification of muons relies mostly on charged-particle penetration in the KLM for momenta larger than 0.7\gevc and on information from the drift chamber and the calorimeter otherwise. 
The selection retains 93\%--99\% of the muons, and rejects 80\%--97\% of the pions, depending on their momenta.
Electrons are identified mostly by comparing measured  momenta  to the energies of the associated calorimeter deposits.
Photons are reconstructed from calorimeter energy deposits greater than 100~MeV that are not associated with any track.
Details of particle reconstruction and identification algorithms are given in Refs.~\cite{ref:b2tip, tracking}.

\subsection{Candidate selections}\label{sec:general}
We require that events have exactly four charged particles with zero net charge and invariant mass \mfourmu\ between 10\gevcc and 11\gevcc.
To suppress backgrounds from misreconstructed and single-beam induced tracks, the transverse and longitudinal projections of the distance of closest approach to the  interaction point of the tracks must be smaller than 0.5\,cm and 2.0\,cm, respectively.
At least three of the tracks must be identified as muons.
This requirement provides better performance than requiring four identified muons or a pair of same-sign muons.
It rejects almost all backgrounds other than \fourmu, while retaining good efficiency for signal. 

In the low dimuon-mass region below 1\gevcc, there are residual backgrounds from $e^+e^-\to\mu^+\mu^-\gamma$, in which the photon converts to an electron-positron pair, and \eemumu\ events. Some of these electrons that are misidentified as muons have low momenta, and thus do not reach the KLM.
The remaining electrons leave signals in the KLM at the gap between the barrel and end cap or in the gaps between adjacent modules. 
In this mass region, we therefore require that no track be identified as an electron. 
 
To suppress radiative backgrounds and, in general, backgrounds with neutral particles, we require that the total energy of all photons be less than 0.4\gev.  

We add a further requirement when $M({{4\mu}})<10.4\gevcc$, exploiting the correlation between the invariant mass and the initial state radiation. This additional selection requires the total energy of all photons to be less than the expected energy of a single irradiated photon, which depends linearly on $M({{4\mu}})$.

In addition, we reject events in which the angle in the c.m.\ frame between the momentum of the  four-muon system and that of the system composed of all the photons is larger than $160^\circ$. 

\begin{figure}
  \begin{center}
    \includegraphics[width=\linewidth]{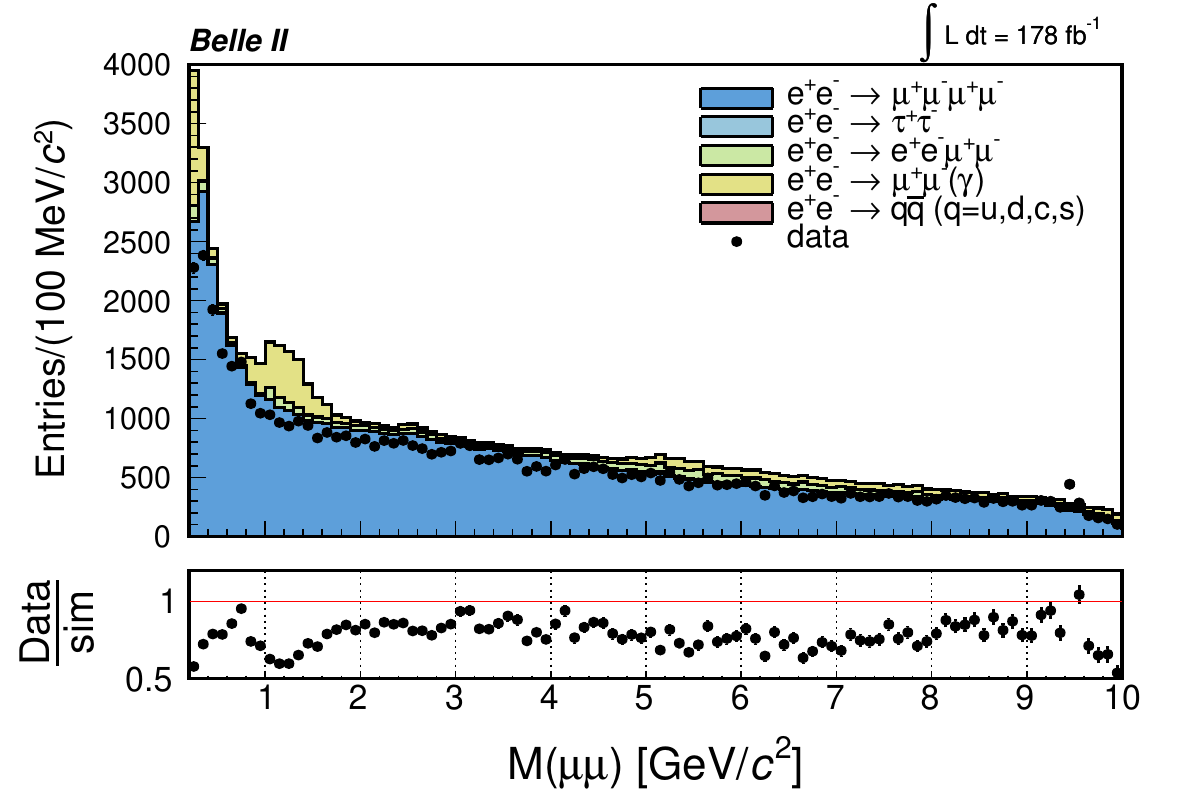}
    \caption{Dimuon invariant-mass distribution in data and simulation for candidates passing all selections but the final background suppression. Contributions from the various simulated processes are stacked. The subpanel shows the data-to-simulation ratio.}
    \label{fig:FU_Mmumu_preMLP}
  \end{center}
\end{figure}

At this level of the analysis, there is no a priori attempt to select a single 
$\mu^+\mu^-$ pair as a candidate $X$ decay.
Each event includes four possible $\mu^+\mu^-$ candidates, each with a different dimuon mass \Mmumu, causing some combinatorial background. 
For each $\mu^+\mu^-$ candidate, the pair of the two remaining muons is labeled as the ``recoil'' pair. 
We consider independently all the $\mu^+\mu^-$ candidates, each with its recoil muons.

The resulting candidate \Mmumu\ distribution is shown in Fig.~\ref{fig:FU_Mmumu_preMLP}. 
The average data-to-simulation yield ratio is 0.76, due to the lack of ISR in the \texttt{AAFH} four-muon generator, in agreement with the values previously reported by \textit{BABAR}~\cite{TheBABAR:2016rlg} and Belle~\cite{PhysRevD.106.012003}.
The excess of the simulation over data in the mass region below 2\gevcc\ is also due to an overestimate of the three-track-trigger efficiency for very low transverse-momentum tracks. Specifically, the enhancement in the range 1--2\gevcc\ originates from the process $e^+e^- \to \mu^+\mu^- \gamma$ with a near-beam-energy photon, followed by conversion of the photon into electron-positron pairs in detector material.
These events are almost entirely removed by the final background suppression. 
Other visible features include the unsimulated contributions from the $\rho$, $J/\psi$, and $\Upsilon(1S)$ resonances.

\subsection{Final background suppression}\label{sec:MLP}
The final selection relies on a few distinctive features that allow the discrimination of signal from background: signal events include a \mumu resonance, which can be seen both in the candidate muon pair and in the  mass of the system recoiling against the two recoil muons; 
the signal is emitted through FSR from a muon (Fig.~\ref{fig:Zp_diagram}), while the dominant four-muon background proceeds through 
double-photon-conversion process (Fig.~\ref{fig:bkg_diag}, left); and the double-photon-conversion process has a distinctive momentum distribution. In the following, some of the relevant variables sensitive to these three classes of features are discussed: they are based both on the \mumu candidate, where we search for signal, and on the recoil muons. For illustration, we show the case for a \zprime\ signal with $m_{Z^\prime}=3\gevcc$ and for background, both with reconstructed candidate dimuon masses $2.75 < M(\mu\mu)< 3.25\gevcc$. The background in this mass region is dominated by the \fourmu\ process, see Fig. \ref{fig:FU_Mmumu_preMLP}.

\begin{figure*}[hbt!]
\centering
\includegraphics[width=0.49\linewidth]
{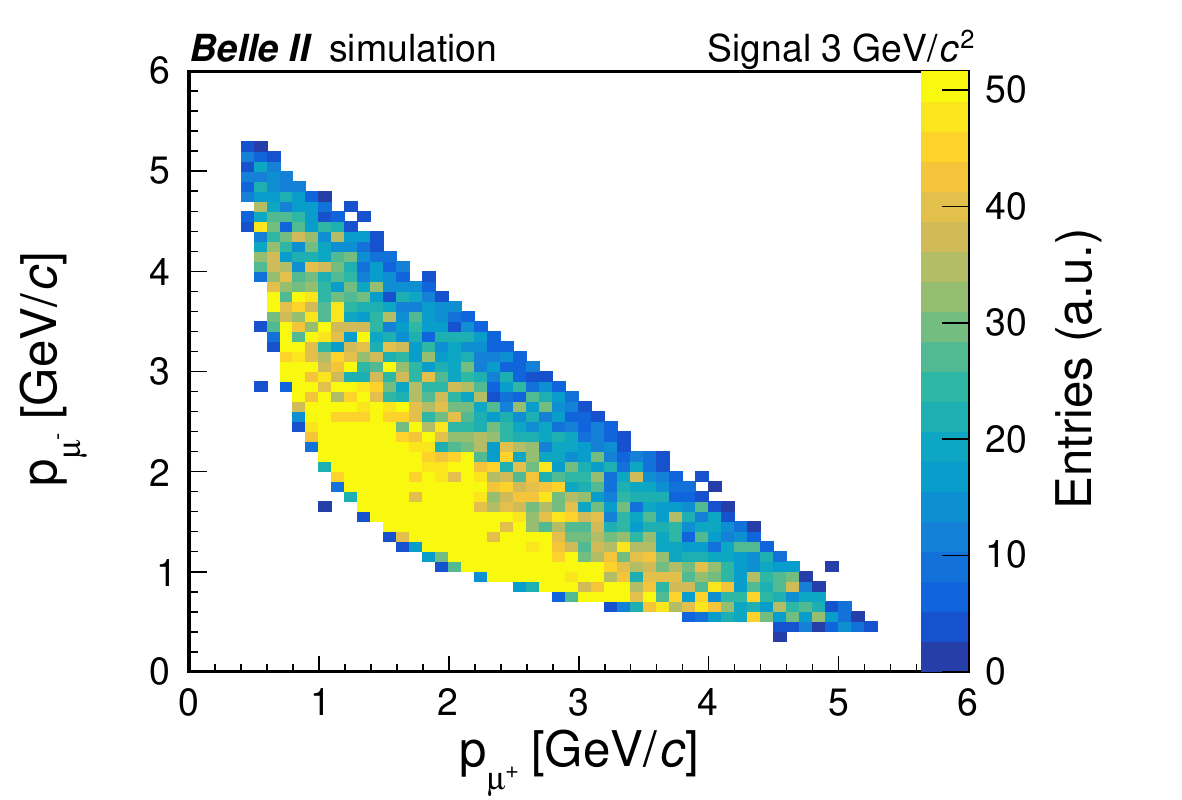}
\includegraphics[width=0.49\linewidth]
{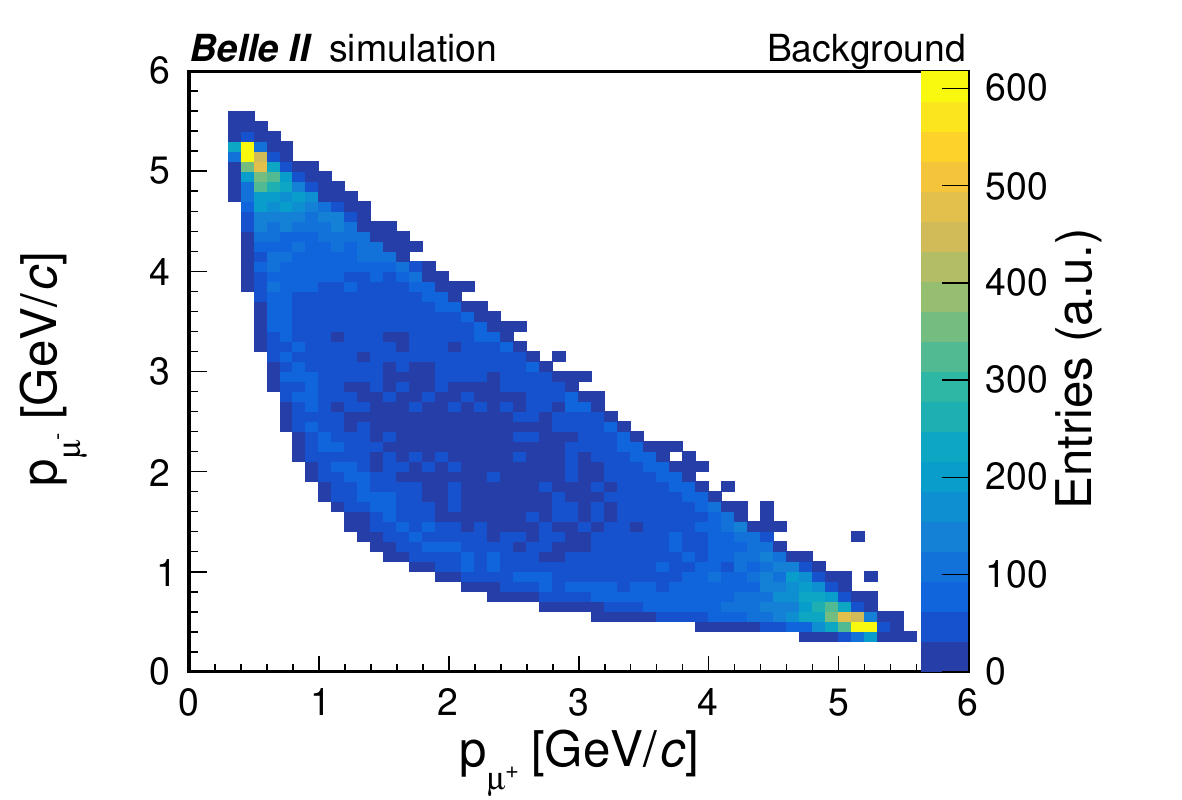}
\caption{\label{fig:Zp_p0p1}Candidate-$\mu^+$ momentum versus candidate-$\mu^-$ momentum for simulated signal (left) with $m_{Z^\prime}=3\gevcc$ and simulated background (right), for dimuon masses  $2.75 < M(\mu\mu)< 3.25\gevcc$.}
\end{figure*}
\begin{figure*}[hbt!]
\centering
\includegraphics[width=0.49\linewidth]
{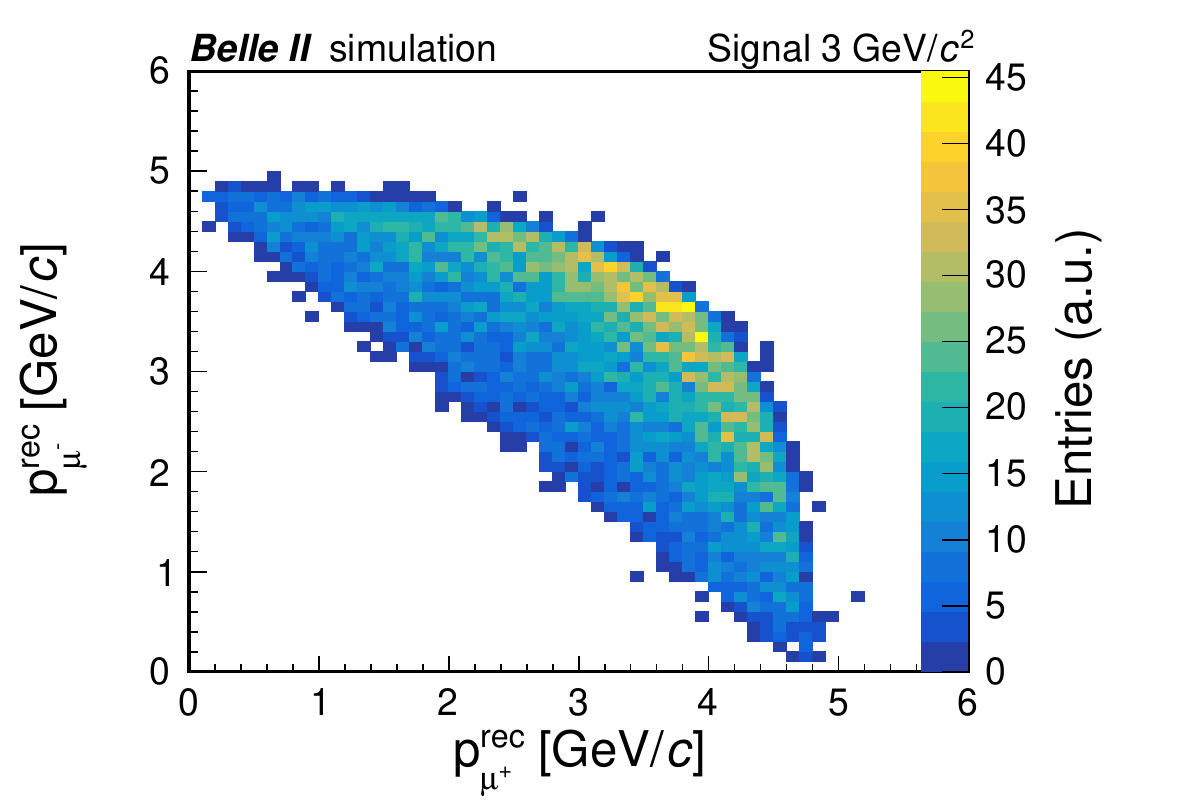}
\includegraphics[width=0.49\linewidth]
{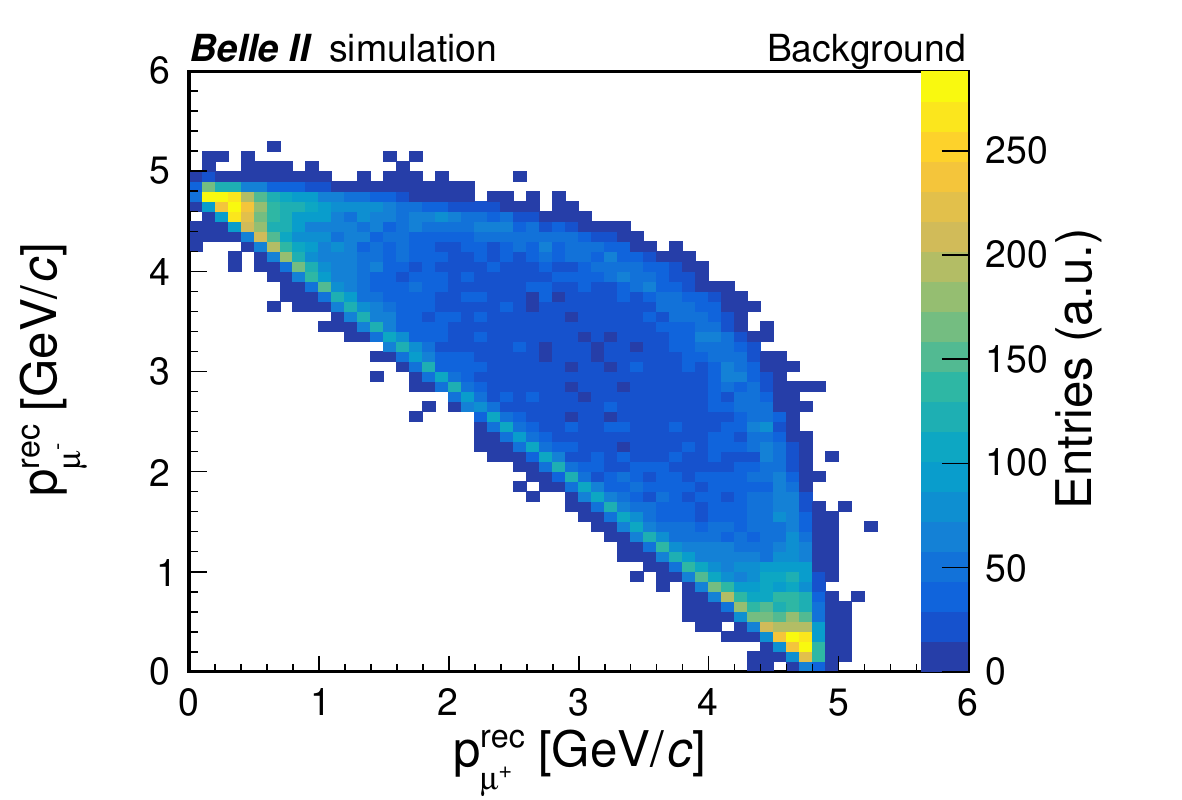}
\caption{\label{fig:Zp_p0p1rec}Recoil-$\mu^+$ momentum versus recoil-$\mu^-$ momentum for simulated signal (left) with $m_{Z^\prime}=3\gevcc$ and simulated  background (right), for dimuon masses  $2.75 < M(\mu\mu)< 3.25\gevcc$.}
\end{figure*}
\begin{figure*}[hbt!]
\centering
\includegraphics[width=0.49\linewidth]{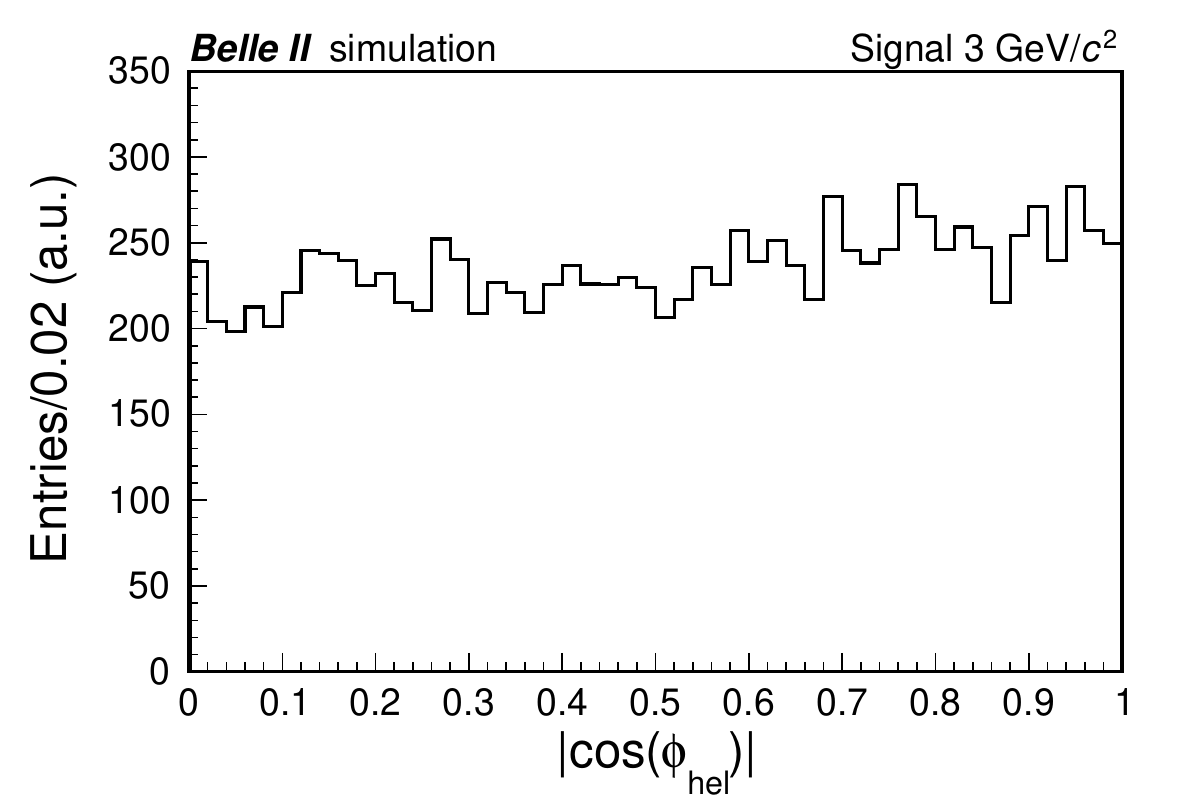}
\includegraphics[width=0.49\linewidth]{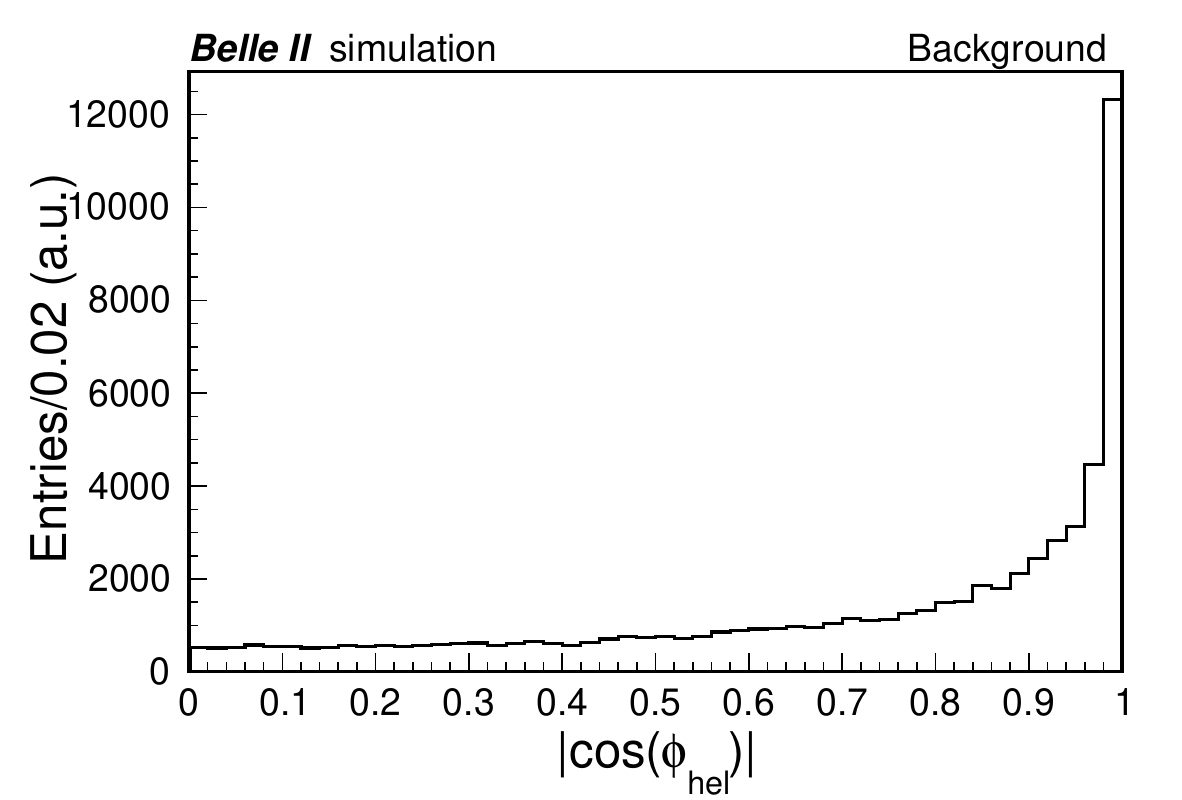}
\caption{\label{fig:Zp_hel}Absolute value of the cosine of the helicity angle for simulated signal (left) with $m_{Z^\prime}=3\gevcc$ and simulated background (right), for dimuon masses $2.75 < M(\mu\mu)< 3.25\gevcc$.} 
\end{figure*}
\begin{figure*}[hbt!]
\centering
\includegraphics[width=0.49\linewidth]
{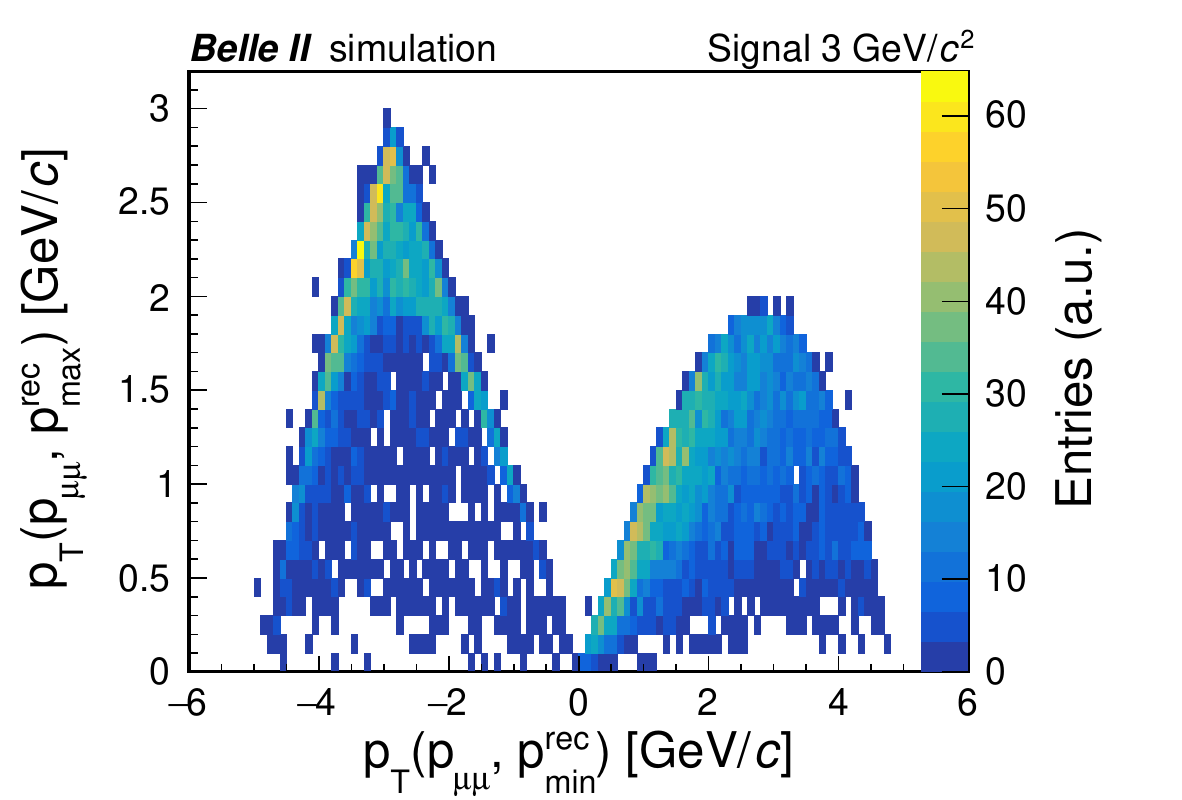}
\includegraphics[width=0.49\linewidth]
{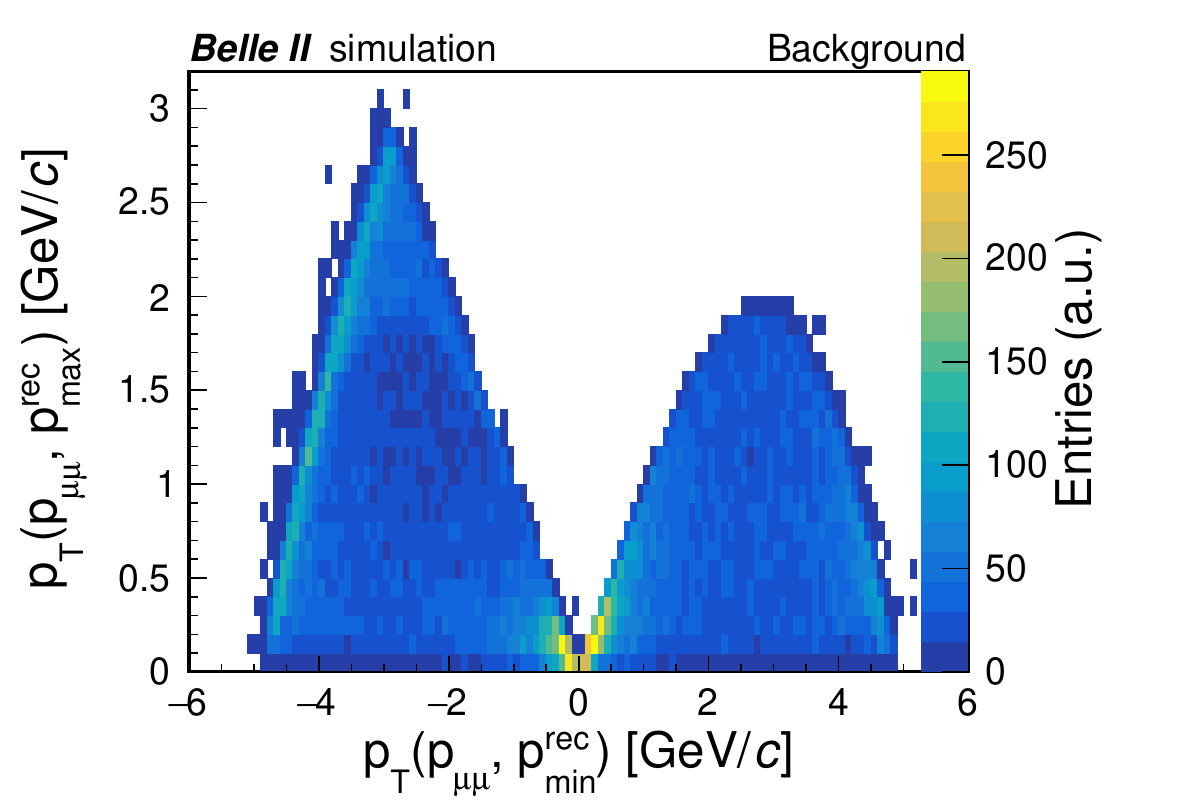}
\caption{ \label{fig:Zp_ptminmax}Candidate-muon-pair transverse momentum with respect to the maximum momentum recoil-muon direction versus the candidate-muon-pair transverse momentum with respect to the minimum momentum recoil-muon direction (with the sign of the longitudinal projection) for simulated signal (left) with $m_{Z^\prime}=3\gevcc$ and simulated background (right), for dimuon masses $2.75 < M(\mu\mu)< 3.25\gevcc$.} 
\end{figure*}
\begin{figure*}[hbt!]
\centering
\includegraphics[width=0.49\linewidth]
{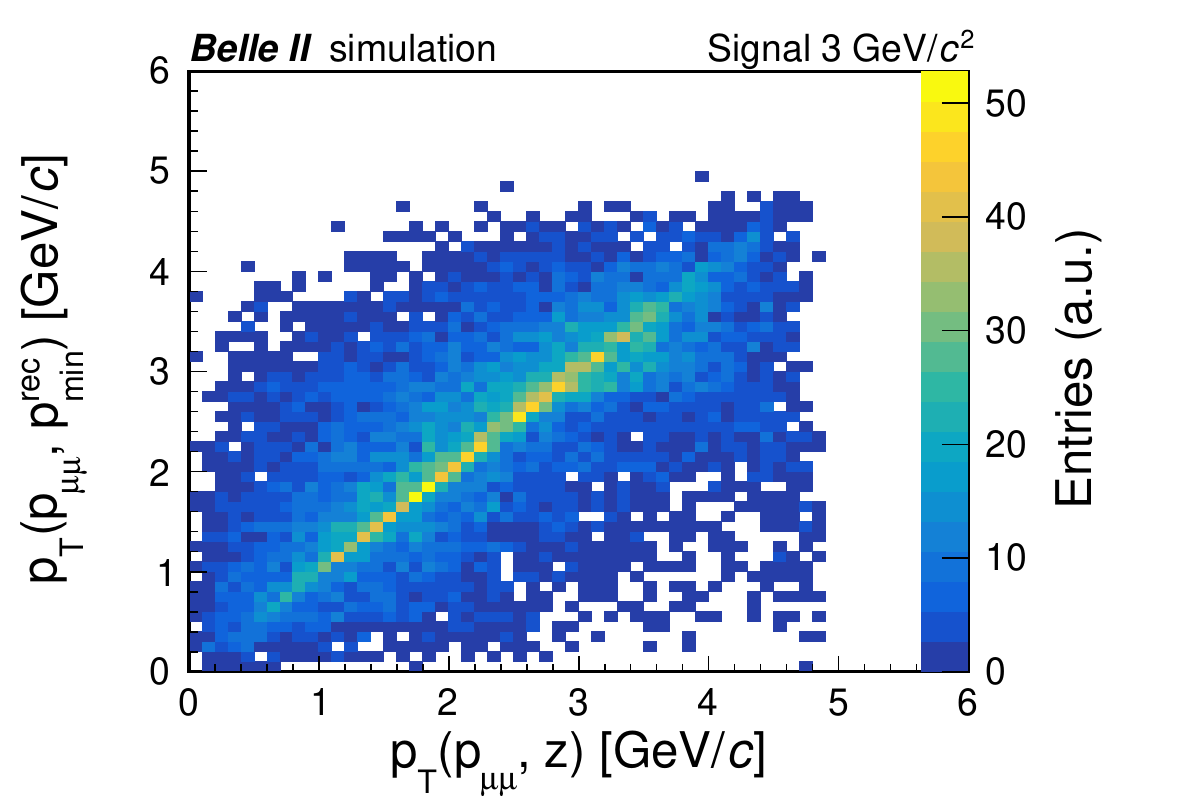}
\includegraphics[width=0.49\linewidth]
{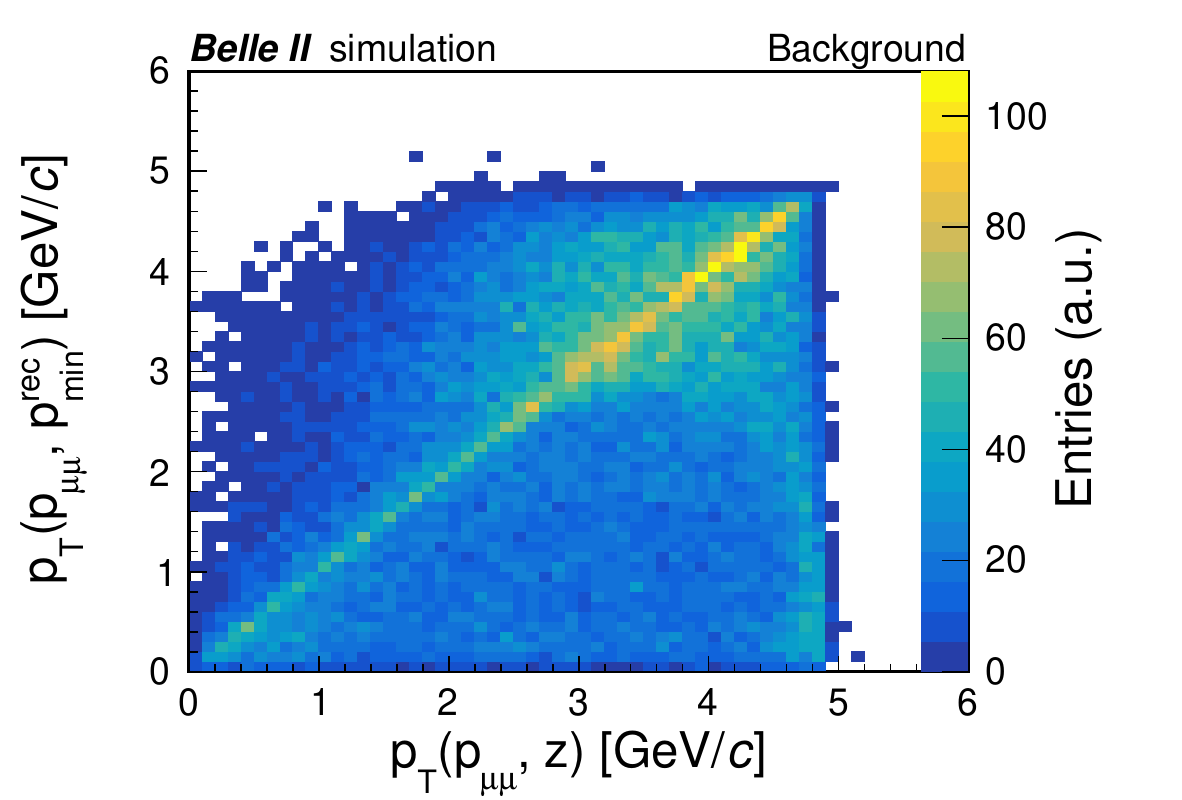}
\caption{\label{fig:Zp_p-ptmin}Candidate-muon-pair transverse momentum with respect to the minimum-momentum recoil-muon direction versus candidate-muon-pair transverse momentum with respect to the beam direction, for simulated signal (left) with $m_{Z^\prime}=3\gevcc$ and simulated background (right), for dimuon masses $2.75 < M(\mu\mu)< 3.25\gevcc$.}
\end{figure*}

Magnitudes of the two candidate muon momenta, $p_{\mu^+}$ and $p_{\mu^-}$, and their correlations are sensitive to the presence of a resonance (Fig.~\ref{fig:Zp_p0p1}). Signal events cluster preferentially in the central part of the distribution, while background predominantly populates the extremes.
A similar effect occurs for the momenta of the two recoil muons,  $p^{\rm rec}_{\mu^+}$ and $p^{\rm rec}_{\mu^-}$ (Fig.~\ref{fig:Zp_p0p1rec}), 
which provide instrumentally uncorrelated access to the same information, though with a different resolution.
The cosine of the helicity angle of the candidate-muon pair $\cos\phi_{\rm hel}$, defined as the angle between the momentum direction of the c.m.\ frame and the $\mu^-$ in the candidate-muon-pair frame, has a uniform distribution for a scalar or an unpolarized massive vector decaying to two fermions, but not for the background processes (Fig.~\ref{fig:Zp_hel}). The slight departure from uniformity in the signal case is due to  momentum resolution, which smears the determination of the boost to the muon-pair frame. 

The double-photon-conversion process (Fig.~\ref{fig:bkg_diag}, left) accounts for 80\% of the four-muon background cross section.
It also includes the case of off-shell photon emission (and subsequent dimuon production) from one of the initial-state electrons, ISR double-photon conversion, which contributes mainly in the low mass region. The annihilation process (Fig.~\ref{fig:bkg_diag}, right) is very similar to  the signal process and constitutes an nearly irreducible background: it accounts for 20\% of the cross section for \Mmumu$< $1\gevcc\ and for 10\% above.
Transverse projections of the candidate-muon-pair momentum $p_{\mu\mu}$ on the direction of the recoil muon with minimum momentum, $p_T(p_{\mu\mu},p^{\rm rec}_{\rm min})$, and on the direction of the recoil muon with maximum momentum, $p_T(p_{\mu\mu},p^{\rm rec}_{\rm max})$, are sensitive to FSR emission (Fig.~\ref{fig:Zp_ptminmax}).
This is because, in case of signal, these are the transverse momenta of $X$ with respect to the direction of the muon from which it was emitted, and with respect to the direction of the other muon. We assign to the transverse projection $p_T(p_{\mu\mu},p^{\rm rec}_{\rm min})$ the sign of the longitudinal projection, since this slightly increases the discriminating power.  
The transverse momentum of the candidate muon pair with respect to the $z$ axis, $p_T(p_{\mu\mu}, z)$, which approximates the beam direction, 
is sensitive to the ISR double-photon conversion mechanism of emission because $p_T(p_{\mu\mu}, z)$ is the transverse momentum of the muon pair with respect to the initial-state-electron direction.
This variable is shown in Fig.~\ref{fig:Zp_p-ptmin} in a two-dimensional distribution versus $p_T(p_{\mu\mu},p^{\rm rec}_{\rm min})$ to illustrate the correlation between variables sensitive to ISR and FSR, respectively. 

\begin{figure*}
\centering
\includegraphics[width=0.49\linewidth]{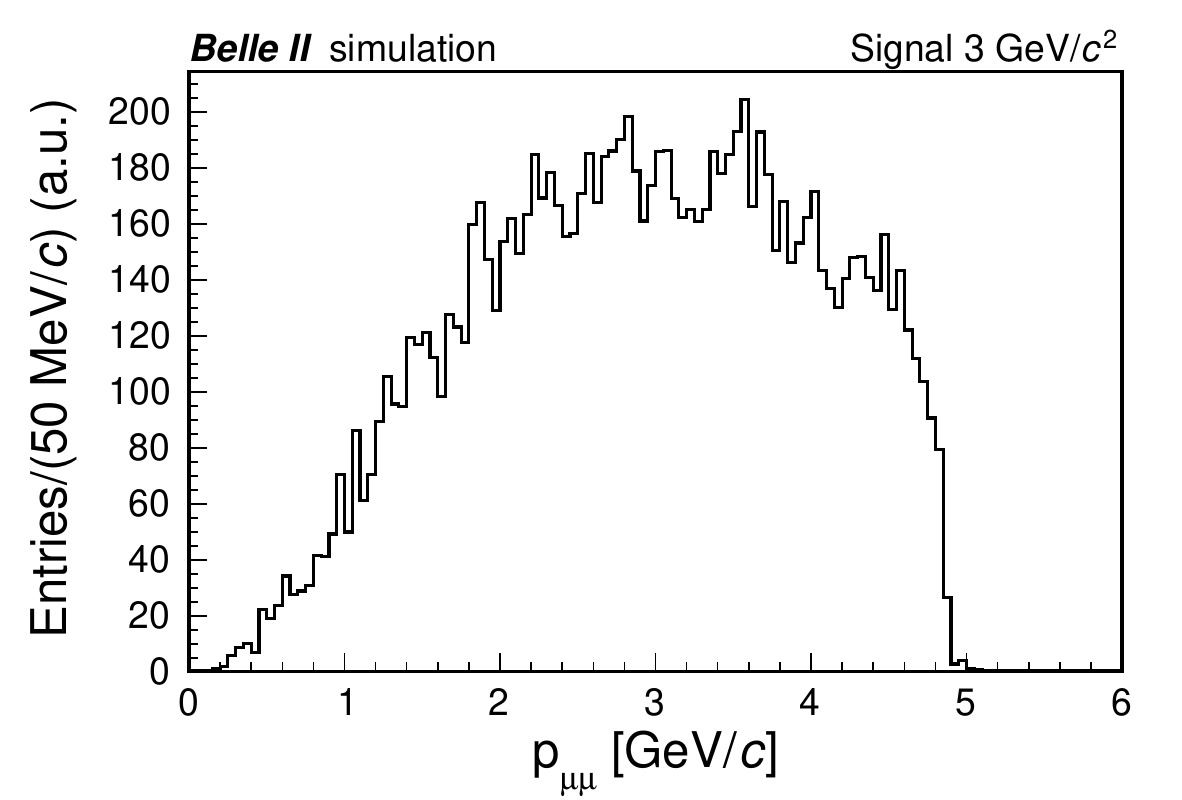}
\includegraphics[width=0.49\linewidth]{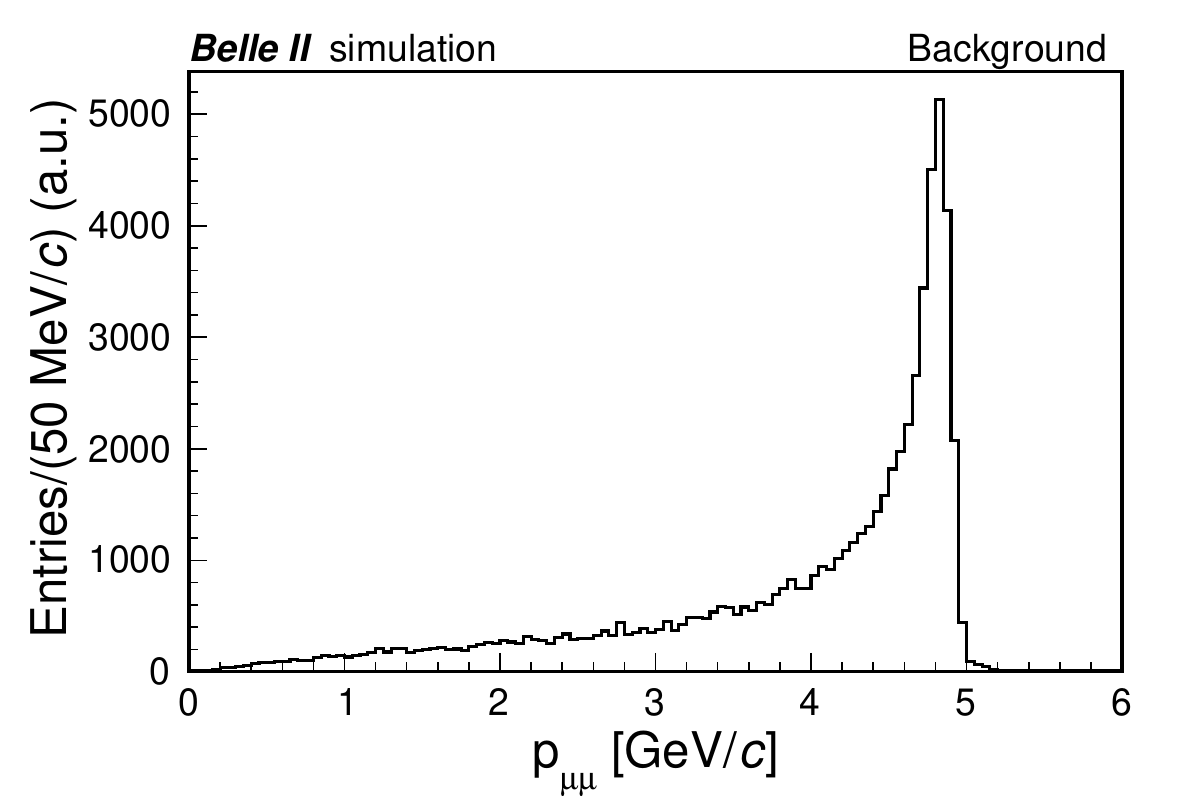}
\caption{\label{fig:Zp_P}Candidate-muon-pair momentum $p_{\mu\mu}$ for signal (left) with $m_{Z^\prime}=3\gevcc$ and background (right), for dimuon masses $2.75 < M(\mu\mu)< 3.25\gevcc$.} 
\end{figure*}

The double photon conversion process produces two muon pairs from two off-shell photons. 
The dominant background at a mass $m_0$ is produced when one pair has \Mmumu\ near $m_0$ and the other pair has a mass at the lowest possible value above $2m_{\mu}$. In these cases, the c.m.\ momentum $p_0$ of the two pairs can be analytically calculated. 
In \fourmu\ background events the dimuon c.m.\ momentum $p_{\mu\mu}$ peaks at $p_0$, in contrast to the signal, at least for two of the  dimuon candidates. This difference is visible in Fig.~\ref{fig:Zp_P}. 

We select sixteen discriminating variables:  the magnitude of the candidate-muon-pair momentum ${p}_{\mu\mu}$; the absolute value of the cosine of the helicity angle in the candidate-muon-pair rest frame; the magnitudes of the candidate-single-muon momenta; the  candidate-single-muon transverse momenta;  the magnitudes of the recoil-single-muon momenta; the recoil-single-muon transverse momenta; $p_T(p_{\mu\mu},p^{\rm rec}_{\rm min})$ and $p_T(p_{\mu\mu},p^{\rm rec}_{\rm max})$; the correlation of $p_T(p_{\mu\mu},p^{\rm rec}_{\rm min})$
with $p_T(p_{\mu\mu}, z)$; and the transverse projections of the recoil-muon-pair momentum on the directions of the momenta of the candidate muons with minimum and maximum momentum. 
All variables other than the helicity angle are defined in the c.m.\ frame. 

We use multilayer perceptron (MLP) artificial neural networks~\cite{hoecker2009tmva} with 16 input neurons, fed with the discriminant variables, and with one  output neuron.
The MLPs are developed using simulated \zprime\ and simulated background events. 
To improve performance, we use five separate MLPs in different \Mmumu\ intervals, which we refer to as MLP ranges: 0.21--1.00\gevcc, 1.00--3.75\gevcc, 3.75--6.25\gevcc, 6.25--8.25\gevcc, and 8.25--10.00\gevcc.
Within the MLP ranges, better performances are obtained if the dependence of the input variables on \zprimemass\ is reduced. This is achieved by scaling the momentum-dimensioned variables by $p_0$, which is the maximum c.m. momentum of the two muon pairs.
To ensure that MLPs are not biased to specific mass values, we use a training signal sample that has mass steps of 5\mevcc, so as to approximate a continuous distribution.
For nearly all masses, the most discriminating variable is $p_{\mu\mu}$, followed by the correlation of $p_{\mu^+}$ and $p_{\mu^-}$. 

The selection applied on the MLP output is studied separately in each MLP range, by maximizing the figure of merit described in Ref.~\cite{Punzi:2003bu}, and then expressed as a function of \Mmumu\ by interpolation. 
The background rejection factor achieved by the MLP selection varies from  2.5 to 14, with the best value around 5\gevcc. 
The resulting background is composed almost entirely of \fourmu\ events, with \mumugamma\ and \eemumu\ processes contributing only below 1\gevcc.
The MLP selection is applied separately to each of the four candidates per event, reducing the average candidate multiplicity per background event to 1.7. 
The average candidate multiplicity per signal event varies between 1.4 and 3, depending on the mass.

\begin{figure}
\begin{center}
\includegraphics[width=0.98\linewidth]{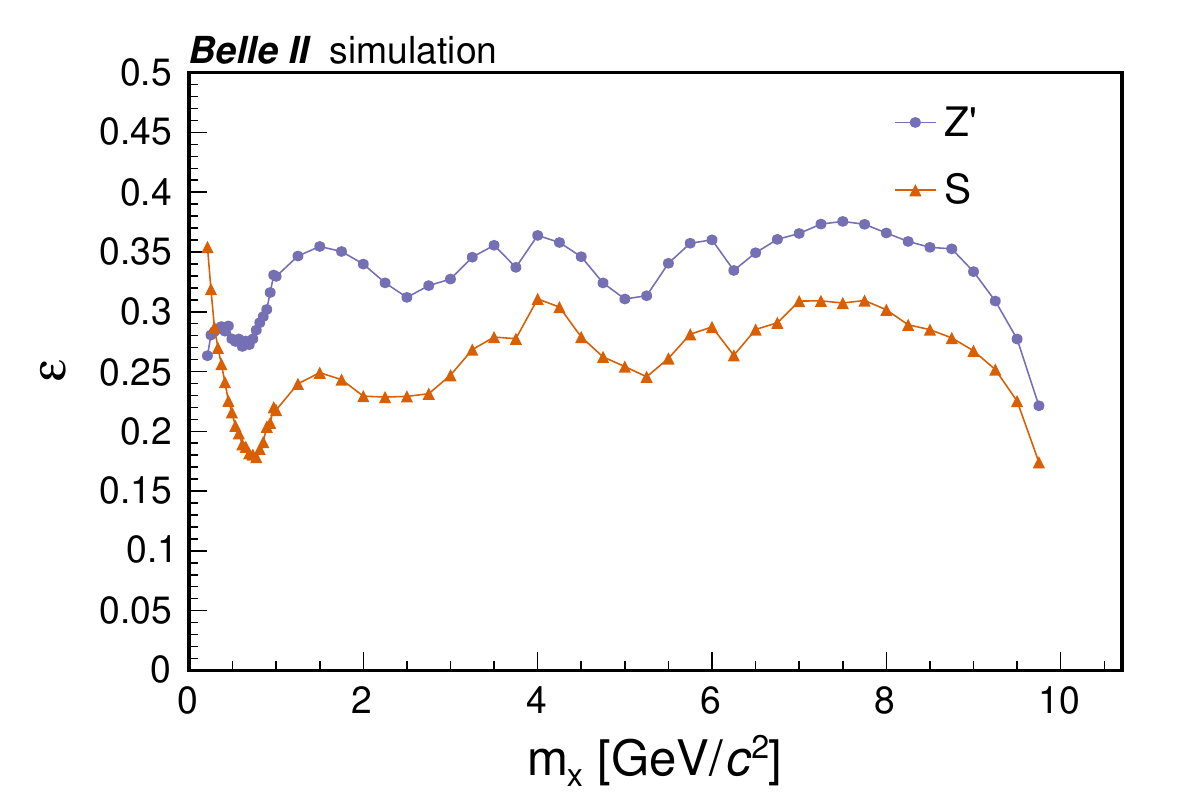}
\end{center}
\caption{\label{fig:efficiency}Signal efficiency as a function of $m_{Z^\prime}$ (purple dots) and $m_S$ (orange triangles) masses after all selections are applied.}
\end{figure}

\subsection{Efficiencies and dimuon spectrum}\label{sec:efficiency}
The efficiencies of the full selection for the \lmultau\ and muonphilic scalar models are shown in Fig.~\ref{fig:efficiency}.
The $S$ boson, due to angular momentum conservation, is produced through a p-wave process, and has a 
higher momentum spectrum than the \zprime, which is produced via an s-wave process. For masses below 1\gevcc, this implies the presence of higher momentum muons in the case of the scalar, which are better identified and detected with higher efficiency.
For masses above 1\gevcc, the muon identification efficiencies for $S$ and \zprime\ are similar, and the higher signal efficiencies for the \zprime\ are due to the differences in the distributions of the momentum-dimensioned input variables, with the MLP optimized for the \lmultau\ model.
The signal efficiencies shown here are corrected for ISR.  
Although the signal generator includes ISR, it does not include the large-angle hard-radiation component that can produce photons in the acceptance, and thereby veto events.
This effect is studied using $e^+e^- \to \mu^+ \mu^- \gamma$  events, generated with \texttt{KKMC} that simulates ISR in a complete way. We require the dimuon mass in the range 10--11\gevcc, to emulate the selection we apply on \mfourmu, that intrinsically limits the maximum energy at which a photon can be radiated. Applying the selection on the photons (see Sec.~\ref{sec:selections}) gives a relative reduction of 2.8\% in efficiency.
\begin{figure}
\centering
\includegraphics[width=0.98\linewidth]{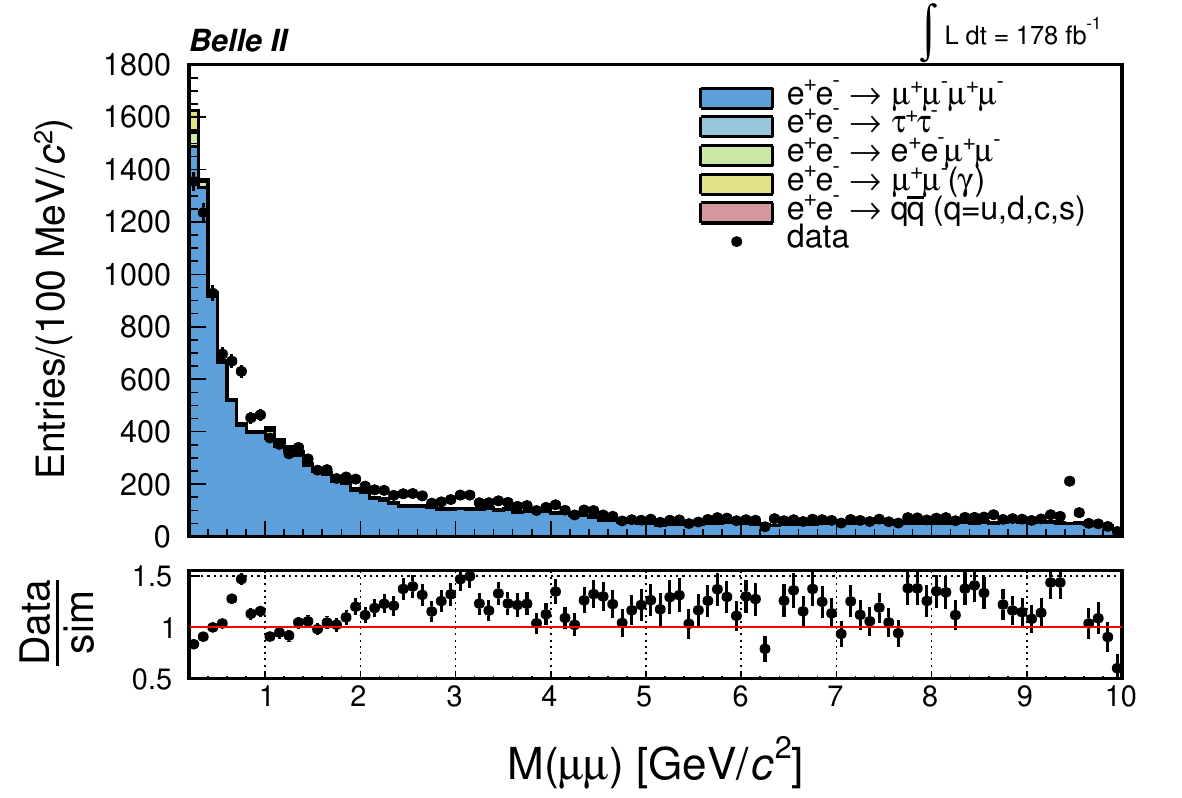}
\caption{\label{fig:M_mumu_afterMLP}Dimuon invariant-mass distribution in data and simulation for candidates passing all the selections. Contributions from the various simulated process are stacked. The subpanel shows the data-to-simulation ratio.}
\end{figure}
To improve the \xmass\ resolution, a kinematic fit is applied requiring that the sum of the four-momenta of the muons be equal to the four-momentum of the c.m.\ system, thereby constraining the four-muon invariant mass to $\sqrt{s}/c^2$. The resulting \Mmumu\ distribution is shown in Fig.~\ref{fig:M_mumu_afterMLP}. 
With the exception of the very low mass region, the data-to-simulation yield ratio is generally above one.
This is because the MLPs perform worse on data, which naturally includes ISR, than on background simulation, which does not.
This is not the case for the signal, which is simulated with the ISR contribution.
Also visible in Fig.~\ref{fig:M_mumu_afterMLP} are modulations originating from the five MLP ranges.
Neither of these effects produce narrow peaking structures at the scale of the signal resolution, 2--5\mevcc (Sec.~\ref{sec:fit}).
As in Fig.~\ref{fig:FU_Mmumu_preMLP}, contributions from the unsimulated $\rho$, $J/\psi$, and $\Upsilon(1S)$ resonances are visible.  

\section{Signal modeling and fit}\label{sec:fit}
To search for the signal, we use the reduced dimuon mass $M_R \equiv \sqrt{M^2(\mu\mu) - 4 m^2_\mu}$, which has smoother behavior than the dimuon mass near the kinematic threshold.  
The reduced-mass resolution is 2--2.5\mevcc\ for \zprimemass\ below 1\gevcc, increases smoothly to 5\mevcc\ for  \zprimemass\ around 5\gevcc, then decreases to 2.5\mevcc\ at 9\gevcc.

The signal yields are obtained from a scan over the $M_R$ spectrum through a series of unbinned maximum likelihood fits.
The signal $M_R$ distributions are parameterized from the simulation as sums of two Crystal Ball functions~\cite{Skwarnicki:1986xj} sharing the same mean.
The background is described with a quadratic function with coefficients as free parameters in the fit for masses below 1\gevcc, and with a straight line above. 
Higher-order polynomials are investigated, but their corresponding fitted coefficients are compatible with zero over the full mass spectrum. 
The broad $\rho$ contribution is accommodated by the quadratic fit.

The scan step-size is set equal to the mass resolution, which is sufficient to detect the presence of a $X$ resonance regardless of its mass.
The fit interval is 60 times the mass resolution, following an optimization study. A total of 2315 fits are performed, covering dimuon masses from 0.212\gevcc to 9\gevcc.
If a fitting interval extends over two different MLP ranges, we use the MLP corresponding to the central mass.   
We exclude the dimuon mass interval 3.07--3.12\gevcc, which corresponds to the $J/\psi$ mass.
The $\Upsilon(1S)$ peak is beyond the mass range of the search.
The fit yields are scaled by 7\% to account for a bias  estimated in a study of the $J/\psi$ in an $e^+e^-\mu^+\mu^-$ control sample, which obtains a width 25\% larger than in simulated signals of that mass.
Propagating this 25\% degradation in resolution to all masses gives an average yield bias of 7\%.
This is also included as a systematic uncertainty (Sec.~\ref{sec:syst}).

Signal yields from the fits are then converted into cross sections, after correcting for signal efficiency and luminosity.

\section{Systematic uncertainties}\label{sec:syst}
Several sources of systematic uncertainties affecting the cross-section determination are taken into account: 
these include signal efficiency, luminosity, and fit procedure.

Uncertainties due to the trigger efficiency in signal events are evaluated by propagating the uncertainties on the measured trigger efficiencies. 
They are 0.3\% for most of the mass spectrum, increasing to 1.7\% at low masses and 0.5\% at high masses.

Uncertainties due to the tracking efficiency are estimated in \tautau\ events in which one $\tau$ decays to a single charged particle and the other $\tau$ to three charged particles.
The relative uncertainty on the signal efficiency is 3.6\%.  

Uncertainties due to the muon identification requirement are studied using $e^+e^- \to \mu^+ \mu^- \gamma$, \eemumu\ events, and final states with a $J/\psi$.
The relative uncertainty on the signal efficiency varies between 0.7\% and 3\%, depending on the $X$ mass. 

Beam backgrounds in the calorimeter can accidentally veto events due to the requirements on photons (Sec.~\ref{sec:general}). 
The effect is studied by changing the level of beam backgrounds in the simulation and by varying the photon energy requirement (see Sec.~\ref{sec:selections}) according to the calorimeter resolution. 
The relative uncertainty on the signal efficiency due to this source is estimated to be below 1\%. 

To evaluate uncertainties due to the data-to-simulation discrepancies in MLP selection efficiencies, 
we  apply a tight selection on \mfourmu\ around $\sqrt{s}/c^2$ requiring it to be in the range 10.54--10.62\gevcc. With this selection, data and background simulation are more directly comparable, because ISR and FSR effects are much less important. 
We compare MLP efficiencies, defined as the ratio of the number of events before and after the MLP selection, in data and simulation and assume that the uncertainties estimated in those signal-like conditions are representative of signal.  
We also assume that these uncertainties hold in the full \mfourmu\ interval 10-11\gevcc for the signal, which is generated with ISR. 
The differences found in each MLP range vary between 1.1\% and 8.1\%, which are taken as estimates of the systematic uncertainties.
To exclude potential bias from the presence of a signal, we check that these differences do not change if we exclude, in each MLP range and for each of the 2315 mass points, intervals ten times larger than the signal mass resolution around the test masses.

Uncertainties due to the interpolation of the signal efficiency between simulated points are estimated to be 3\%, which is assigned as a relative uncertainty on the signal efficiency. 

Uncertainties due to the fit procedure, in addition to that arising from mass resolution, are evaluated using a bootstrap technique~\cite{efron1994introduction}.
A number of simulated signal events corresponding to the yield excluded at 90\% confidence level are overlaid on simulated background and fitted for each \zprime\ mass. The distribution of the difference between the overlaid and the fitted yields, divided by the fit uncertainty, shows a negligible average bias with a width that deviates from one by 4\%, which is assigned as a relative uncertainty on the signal-yield determination.
Additional uncertainties related to the fit procedure are those due to the mass resolution, discussed in Sec.~\ref{sec:fit}. An uncertainty of 7\%, equal to the average yield bias, is included.

Systematic uncertainties from data-to-simulation differences in momentum resolution and beam-energy shift are found to be negligible, due to the kinematic fitting procedure.
Finally, the integrated luminosity has a systematic uncertainty of 1\%~\cite{lumi}. 

The uncertainties are summed in quadrature to give a total that ranges from 9.5\% to 12.9\% depending on the $X$ mass. The contributions to the systematic uncertainty are summarized in Table~\ref{tab:syst-final}.
The systematic uncertainties are included as nuisance parameters with Gaussian constraints on the signal efficiency, with widths equal to the estimated systematic uncertainties.
\begin{table}
    \centering
    \caption{Systematic uncertainties affecting the cross-section determination.}
    \label{tab:syst-final}
    \begin{tabular}{c c c c}
      \hline
      Source          & uncertainty(\%)\\
      \hline
      Trigger     & 0.3--1.7 \\
      Tracking        & 3.6     \\
      Particle identification     & 0.7--3     \\
      Beam background and          & \multirow{2}{*}{1}   \\
      calorimeter energy resolution \\
      MLP selection   & 1.1--8.1      \\
      Efficiency interpolation   & 3       \\
      Fit bias        & 4       \\
      Mass resolution & 7       \\
      Luminosity      & 1       \\
      \hline
      Total & 9.5--12.9 \\
      \hline
    \end{tabular}
\end{table}
\section{Results}\label{sec:results}
\begin{figure}[ht!]
  \begin{center}
    \begin{tabular}{c} 
     \includegraphics[width=0.98\linewidth]{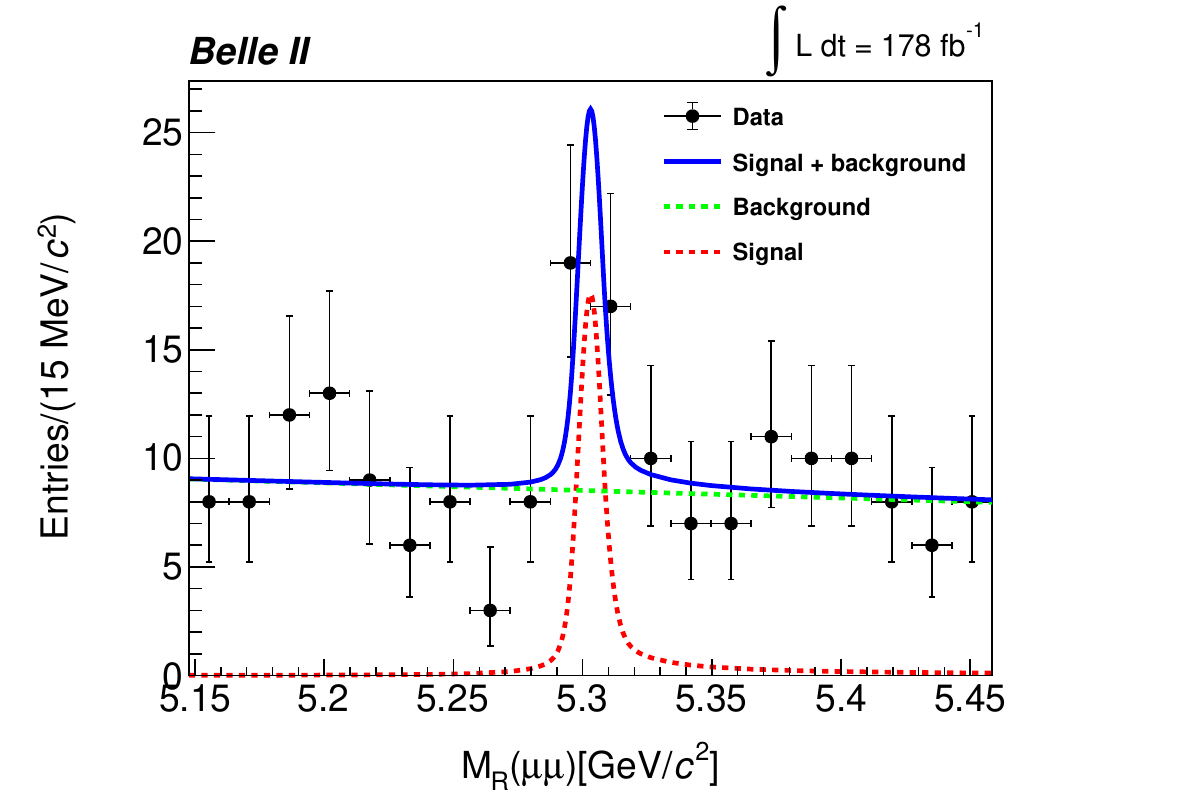}
   \end{tabular}
 \end{center}
 \caption{Fit for a \zprime\ mass hypothesis of 5.307\gevcc, for which we obtain the maximum local significance of $3.4\sigma$.}
 \label{fig:zprimeFit}
\end{figure}

The significance of signal over background for  
each fit is evaluated as $ \sqrt{2\, \text{log} (\mathcal{L}/\mathcal{L}_{0})}$,
where $\mathcal{L}$ and $\mathcal{L}_{0}$ are the likelihoods of the fits with and without signal. 
The largest local one-sided significance observed is 3.4$\sigma$ at $M(\mu\mu) = 5.307\gevcc$, corresponding to a 1.6$\sigma$ global significance after taking into account the look-elsewhere effect~\cite{Cowan:2010js, Gross:2010qma}. The corresponding fit is shown in Fig.~\ref{fig:zprimeFit}. 
Three additional mass points have local significances that  exceed 3$\sigma$.
They are at \Mmumu\ masses of 1.939\gevcc, 4.518\gevcc, and 4.947\gevcc, with global significances of 0.6$\sigma$,  1.2$\sigma$, and 1.1$\sigma$, respectively.

\begin{figure}
  \begin{center}
    \begin{tabular}{c} 
     \includegraphics[width=0.98\linewidth]{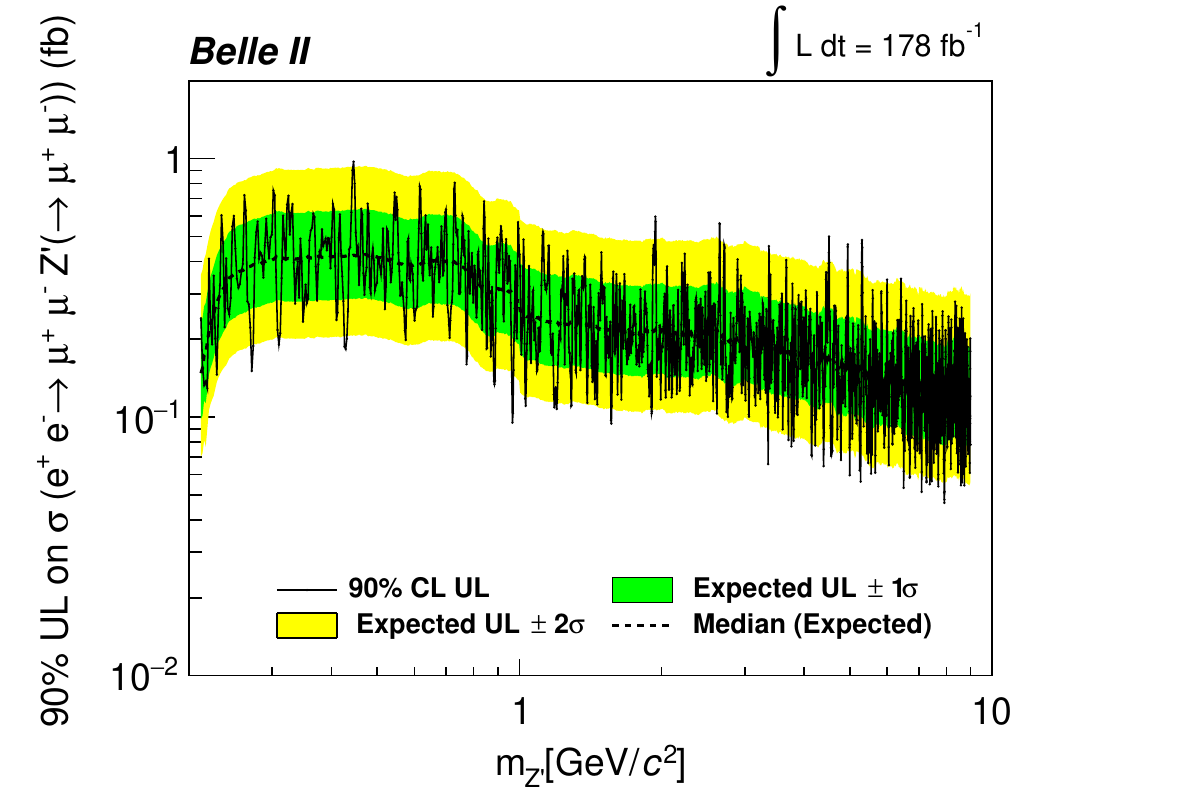}\\
     \includegraphics[width=0.98\linewidth]{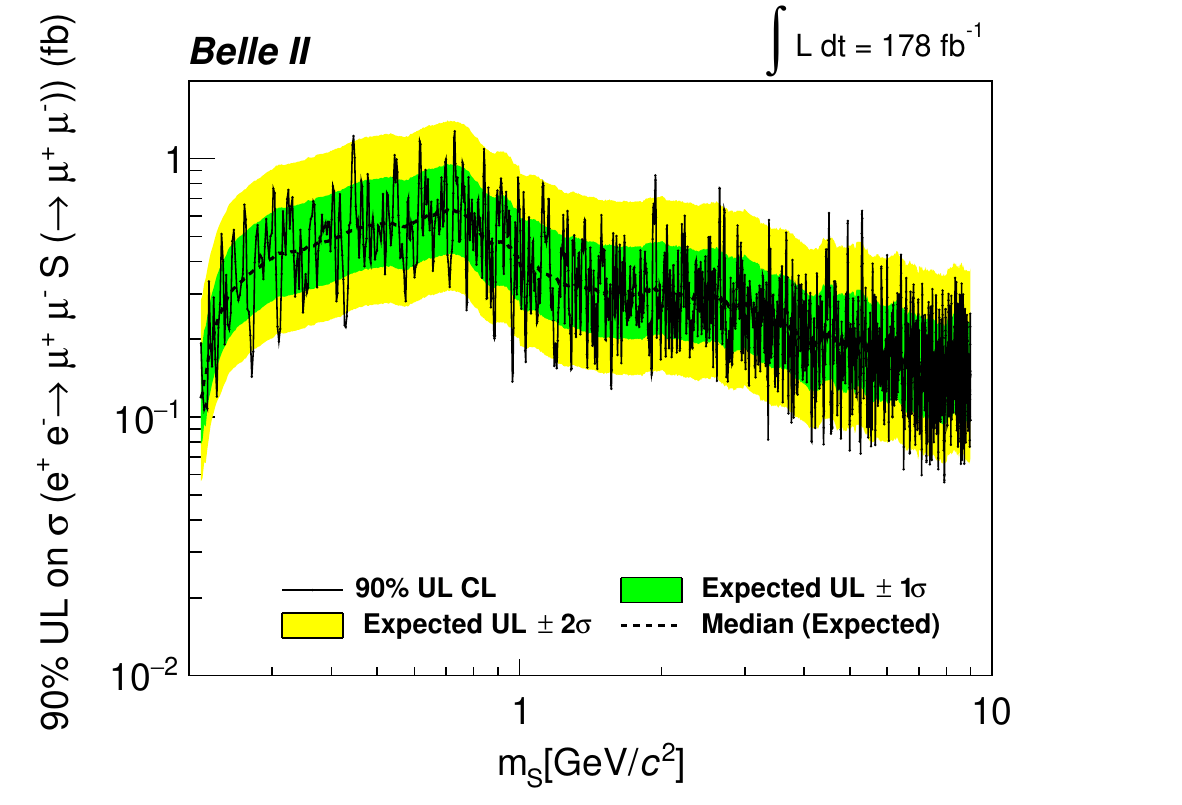}
\end{tabular}
 \end{center}
 \caption{Observed 90\% confidence level upper limits and corresponding expected limits on the cross sections for the processes \Xmumu, as functions of the \zprime\ mass (top) and $S$ mass (bottom).}
 \label{fig:xsec}
\end{figure}

\begin{figure}
  \begin{center}
    \begin{tabular}{c} 
     \includegraphics[width=0.98\linewidth]{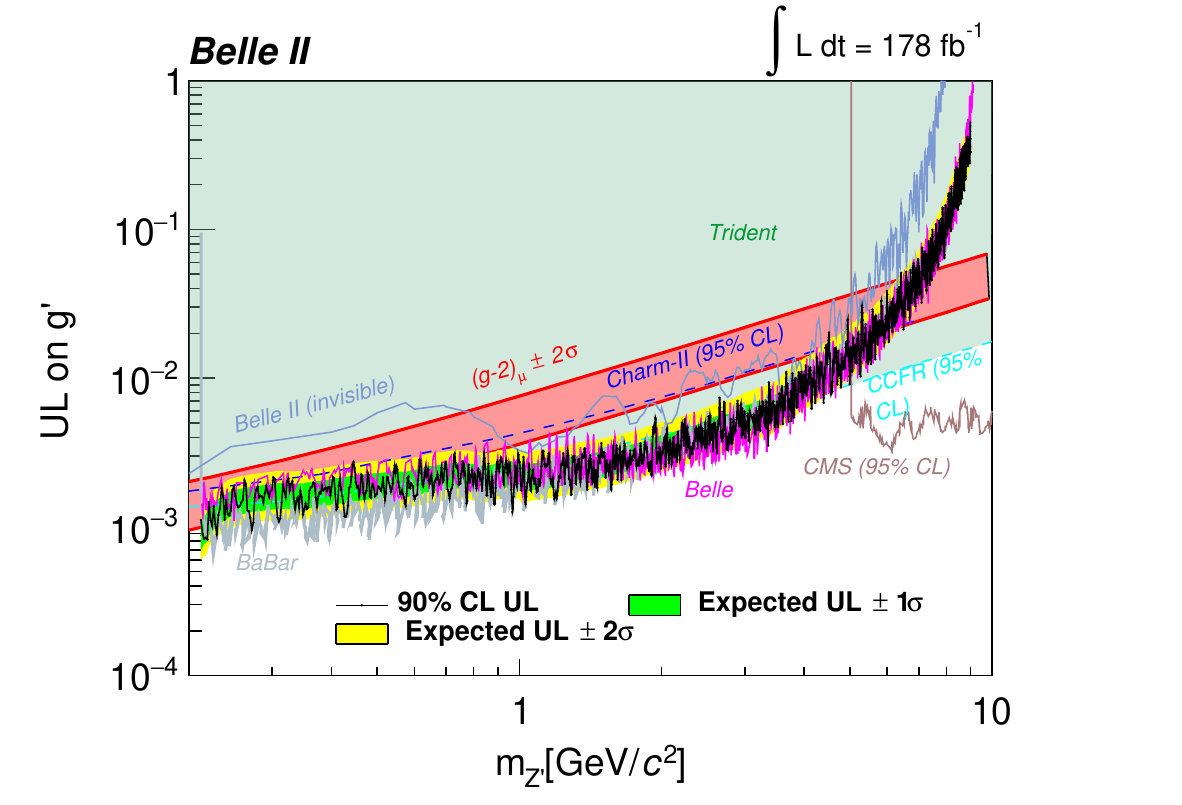}\\
     \includegraphics[width=0.98\linewidth]{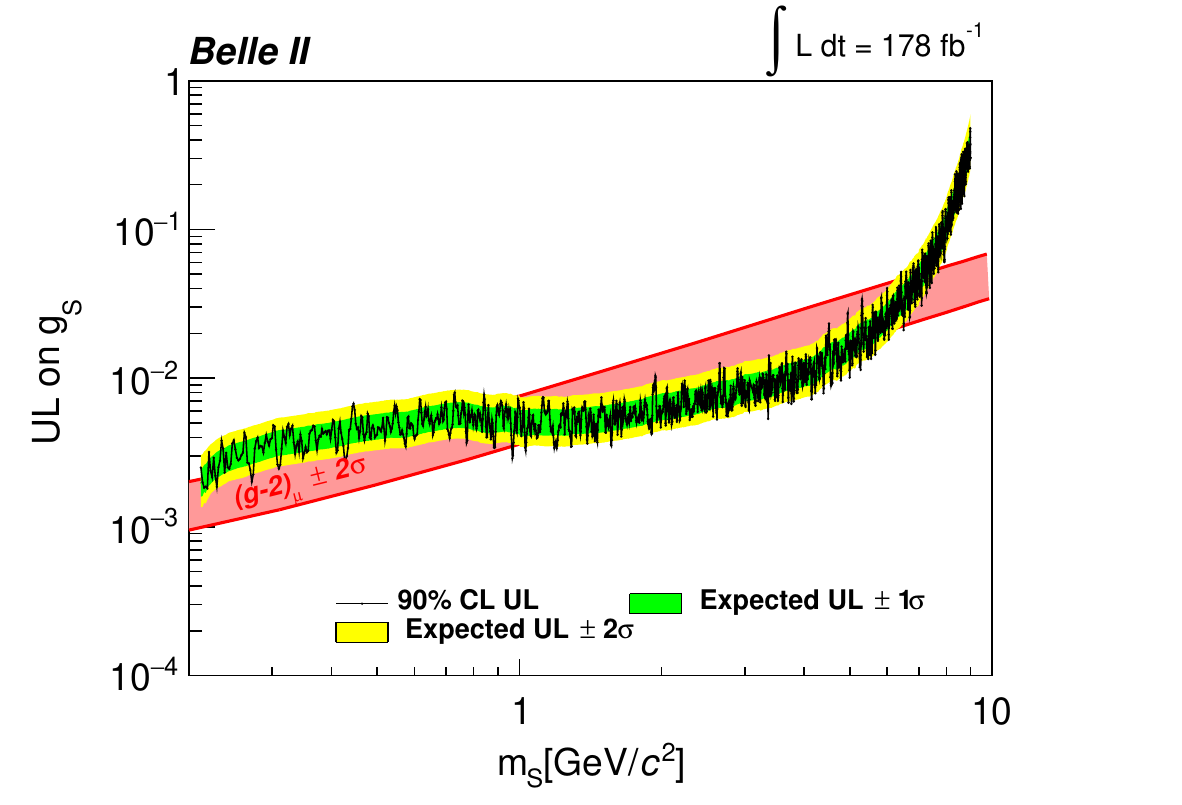}
\end{tabular}
 \end{center}
 \caption{Observed 90\% CL\ upper limits and corresponding expected limits on (top) the \lmultau\ model coupling $g^\prime$ and on (bottom) the muonphilic scalar model coupling  $g_S$.
 Also shown in the top panel are constraints from  Belle~II~\cite{PhysRevLett.124.141801,zpinvis_update} for invisible \zprime\ decays, and from \textit{BABAR}~\cite{TheBABAR:2016rlg}, Belle~\cite{PhysRevD.106.012003}, and CMS~\cite{cms} (95\% CL) searches for \zprime\ decays to muons, along with constraints (95\% CL) derived from the trident production in neutrino experiments~\cite{PhysRevLett.113.091801,PhysRevLett.107.141302,kamada2018self}.
 The red band in each panel shows the region that explains the muon anomalous magnetic moment $(g-2)_{\mu} \pm 2\sigma$.}
 \label{fig:coupling}
\end{figure}

Since we do not observe any significant excess above the background, we derive 90\% confidence level (CL) upper limits (UL) on the process cross sections $\sigma(e^+e^- \to \mu^+\mu^- X) \times \mathcal{B}(X\to \mu^+\mu^-)$ separately for \zprime\ and $S$ (Fig.~\ref{fig:xsec}), using the frequentist procedure $\rm CL_S$~\cite{CLS}. 
The expected limits in Fig.~\ref{fig:xsec} are the median limits from background-only simulated samples that use yields from fits to data.
We obtain upper limits ranging from 0.046\,fb to 0.97\,fb for the \lmultau\ model, and from 0.055\,fb to 1.3\,fb for the muonphilic scalar model.  
These upper limits are dominated by sample size, with systematic uncertainties worsening them on average by less than 1\%.\\
The cross-section results are translated into upper limits on the coupling constant $g^\prime$ of the \lmultau\ model and on the coupling constant $g_S$ of the muonphilic scalar model (Fig.~\ref{fig:coupling}). 
For masses below 6\gevcc, they range from 0.0008 to 0.039 for the \lmultau\ model and from 0.0018 to 0.040  for the muonphilic-scalar model. 
These limits exclude the \lmultau\ model and the muonphilic scalar model as explanations of the $(g-2)_\mu$ anomaly for $0.8<m_{Z^\prime}<4.9$\gevcc and $2.9<m_S<3.5$\gevcc, respectively.\\
Our constraints on $g^\prime$ are similar to those set by \textit{BABAR}~\cite{TheBABAR:2016rlg} for \zprimemass\ above 1\gevcc\ and to those set by Belle~\cite{PhysRevD.106.012003} on the full \zprimemass\ spectrum, both based on much larger integrated luminosities than ours. 
For the muonphilic scalar model, we do not show the constraints in Ref.~\cite{Capdevilla_2022}, since they may not take into account all the experimental details affecting the signal efficiency, particularly those related to the higher momentum spectrum compared to the \zprime.\\
Numerical results of the cross-section and the coupling constant of the \zprime\ and the $S$ are available in the Supplemental Material~\cite{supp}.

\section{Conclusion}\label{sec:summary}

We search for the process \Xmumu\ in a data sample of electron-positron collisions at 10.58\gev collected by Belle~II at SuperKEKB in 2020 and 2021, corresponding to an integrated luminosity of 178\invfb.
We find no significant excess above the background.
We set upper limits on the cross sections for masses between 0.212\gevcc and 9\gevcc, ranging from 0.046\,fb to 0.97\,fb for the \lmultau\ model, and from 0.055\,fb to 1.3\,fb for the muonphilic scalar model. 
We derive exclusion limits on the couplings for the two different models.
For masses below 6\gevcc, they range from 0.0008 to 0.039 for the \lmultau\ model and from 0.0018 to 0.040  for the muonphilic-scalar model. 
These limits exclude the \lmultau\ model and the muonphilic scalar model as explanations of the $(g-2)_\mu$ anomaly for $0.8<m_{Z^\prime}<4.9$\gevcc and $2.9<m_S<3.5$\gevcc, respectively.
These are the first results for the muonphilic scalar model based on a realistic evaluation of the signal efficiency that takes into account all the experimental details.

% Policy from October 20, 2022
This work, based on data collected using the Belle II detector, which was built and commissioned prior to March 2019,
%Belle1 and data collected using the Belle detector, which was operated until June 2010,
was supported by
%Armenia
Higher Education and Science Committee of the Republic of Armenia Grant No.~23LCG-1C011;
%Australia
Australian Research Council and Research Grants
No.~DP200101792, % Jackson
No.~DP210101900, % Urquijo
No.~DP210102831, % Sevior
No.~DE220100462, % Hsu
No.~LE210100098, % Infrastructure
and
No.~LE230100085; % Infrastructure
%Austria
Austrian Federal Ministry of Education, Science and Research,
Austrian Science Fund
No.~P~34529,
No.~J~4731,
No.~J~4625,
and
No.~M~3153,
and
Horizon 2020 ERC Starting Grant No.~947006 ``InterLeptons'';
%Canada
Natural Sciences and Engineering Research Council of Canada, Compute Canada and CANARIE;
%China
National Key R\&D Program of China under Contract No.~2022YFA1601903,
National Natural Science Foundation of China and Research Grants
No.~11575017,
No.~11761141009,
No.~11705209,
No.~11975076,
No.~12135005,
No.~12150004,
No.~12161141008,
and
No.~12175041,
and Shandong Provincial Natural Science Foundation Project~ZR2022JQ02;
%Czech Republic
the Czech Science Foundation Grant No.~22-18469S 
and
Charles University Grant Agency project No.~246122;
%EU
European Research Council, Seventh Framework PIEF-GA-2013-622527,
Horizon 2020 ERC-Advanced Grants No.~267104 and No.~884719,
Horizon 2020 ERC-Consolidator Grant No.~819127,
Horizon 2020 Marie Sklodowska-Curie Grant Agreement No.~700525 ``NIOBE''
and
No.~101026516,
and
Horizon 2020 Marie Sklodowska-Curie RISE project JENNIFER2 Grant Agreement No.~822070 (European grants);
%France
L'Institut National de Physique Nucl\'{e}aire et de Physique des Particules (IN2P3) du CNRS
and
L'Agence Nationale de la Recherche (ANR) under grant No. ANR-21-CE31-0009 (France);
%Germany
BMBF, DFG, HGF, MPG, and AvH Foundation (Germany);
%India
Department of Atomic Energy under Project Identification No.~RTI 4002,
Department of Science and Technology,
and
UPES SEED funding programs
No.~UPES/R\&D-SEED-INFRA/17052023/01 and
No.~UPES/R\&D-SOE/20062022/06 (India);
%Israel
Israel Science Foundation Grant No.~2476/17,
U.S.-Israel Binational Science Foundation Grant No.~2016113, and
Israel Ministry of Science Grant No.~3-16543;
%Italy
Istituto Nazionale di Fisica Nucleare and the Research Grants BELLE2;
%Japan
Japan Society for the Promotion of Science, Grant-in-Aid for Scientific Research Grants
No.~16H03968,
No.~16H03993,
No.~16H06492,
No.~16K05323,
No.~17H01133,
No.~17H05405,
No.~18K03621,
No.~18H03710,
No.~18H05226,
No.~19H00682, % Niigata
No.~20H05850,
No.~20H05858,
No.~22H00144,
No.~22K14056,
No.~22K21347,
No.~23H05433,
No.~26220706,
and
No.~26400255,
%the National Institute of Informatics, and Science Information NETwork 5 (SINET5), 
and
the Ministry of Education, Culture, Sports, Science, and Technology (MEXT) of Japan;  
%Korea
National Research Foundation (NRF) of Korea Grants
No.~2016R1\-D1A1B\-02012900,
No.~2018R1\-A2B\-3003643,
No.~2018R1\-A6A1A\-06024970,
No.~2019R1\-I1A3A\-01058933,
No.~2021R1\-A6A1A\-03043957,
No.~2021R1\-F1A\-1060423,
No.~2021R1\-F1A\-1064008,
No.~2022R1\-A2C\-1003993,
and
No.~RS-2022-00197659,
Radiation Science Research Institute,
Foreign Large-Size Research Facility Application Supporting project,
the Global Science Experimental Data Hub Center of the Korea Institute of Science and Technology Information
and
KREONET/GLORIAD;
%Malaysia
Universiti Malaya RU grant, Akademi Sains Malaysia, and Ministry of Education Malaysia;
%Mexico
% CINVESTAV-IPN, UNAM, UAS, BUAP and CONACYT are funded under
Frontiers of Science Program Contracts
No.~FOINS-296,
No.~CB-221329,
No.~CB-236394,
No.~CB-254409,
and
No.~CB-180023, and SEP-CINVESTAV Research Grant No.~237 (Mexico);
%Poland
the Polish Ministry of Science and Higher Education and the National Science Center;
%Russia
the Ministry of Science and Higher Education of the Russian Federation
and
the HSE University Basic Research Program, Moscow;
%Saudi Arabia
University of Tabuk Research Grants
No.~S-0256-1438 and No.~S-0280-1439 (Saudi Arabia);
%Slovenia
Slovenian Research Agency and Research Grants
No.~J1-9124
and
No.~P1-0135;
%Spain
%Belle1 Ikerbasque, Basque Foundation for Science,
%Belle1 the State Agency for Research of the Spanish Ministry of Science and Innovation through Grant No. PID2022-136510NB-C33,
Agencia Estatal de Investigacion, Spain
Grant No.~RYC2020-029875-I
and
Generalitat Valenciana, Spain
Grant No.~CIDEGENT/2018/020;
%Swiss (Belle 1)
%Belle1 the Swiss National Science Foundation;
%Sweden
The Knut and Alice Wallenberg Foundation (Sweden), Contracts No.~2021.0174 and No.~2021.0299;
%Taiwan
National Science and Technology Council,
and
Ministry of Education (Taiwan);
%Thailand
Thailand Center of Excellence in Physics;
%Turkey
TUBITAK ULAKBIM (Turkey);
%Ukraine
National Research Foundation of Ukraine, Project No.~2020.02/0257,
and
Ministry of Education and Science of Ukraine;
%USA
the U.S. National Science Foundation and Research Grants
No.~PHY-1913789 % Indiana CEEM
and
No.~PHY-2111604, % Luther
and the U.S. Department of Energy and Research Awards
No.~DE-AC06-76RLO1830, % PNNL
No.~DE-SC0007983, % Wayne State
No.~DE-SC0009824, % Florida
No.~DE-SC0009973, % VPI
No.~DE-SC0010007, % Duke
No.~DE-SC0010073, % South Carolina
No.~DE-SC0010118, % Carnegie Mellon
No.~DE-SC0010504, % Hawaii
No.~DE-SC0011784, % Cincinnati
No.~DE-SC0012704, % BNL
No.~DE-SC0019230, % Duke
No.~DE-SC0021274, % Mississippi
No.~DE-SC0021616, % Mississippi
No.~DE-SC0022350, % Louisville
No.~DE-SC0023470; % South Alabama
%last group
and
%Vietnam
the Vietnam Academy of Science and Technology (VAST) under Grants
No.~NVCC.05.12/22-23
and
No.~DL0000.02/24-25.

% Policy from October 20, 2022
These acknowledgements are not to be interpreted as an endorsement of any statement made
by any of our institutes, funding agencies, governments, or their representatives.

We thank the SuperKEKB team for delivering high-luminosity collisions;
the KEK cryogenics group for the efficient operation of the detector solenoid magnet;
the KEK Computer Research Center for on-site computing support; the NII for SINET6 network support;
and the raw-data centers hosted by BNL, DESY, GridKa, IN2P3, INFN, 
%Belle1 PNNL/EMSL, 
and the University of Victoria.

\end{document}

Supplemental material 

\title{Search for a $\mu^+\mu^-$ resonance in four-muon final states at Belle II }

\author{Belle II Collaboration}

\maketitle

\input{supp_content}

%% file: definitions.tex
\usepackage{graphicx} % Include figure files
\usepackage{epstopdf} % Allow to include eps files
\usepackage{dcolumn} % Align table columns on decimal point
\usepackage{xcolor} % Color support
\usepackage{amssymb} % Extended symbol collection
\usepackage{hyperref} % Generate pdfs with hyperlinks
\usepackage[utf8]{inputenc} % Allow UTF8 characters
\usepackage[T1]{fontenc}
\usepackage[english]{babel} % All publications are in English
\usepackage[displaymath, mathlines]{lineno} % Line numbers for drafts
\usepackage{blindtext} % For testing
\usepackage{amsmath}
\usepackage{siunitx}
\usepackage{orcidlink}
\usepackage{colordvi}
\usepackage{color}
\usepackage{amssymb}
\usepackage{url}
\graphicspath{{figures}}
\usepackage{tabularx}
\usepackage{comment}
\usepackage{multirow}

\graphicspath{{figures/}} % Use figures directory for figures

\newcommand {\mumugamma} {$e^{+}e^{-}\rightarrow \mu^+ \mu^- (\gamma)$}
\newcommand {\pipigamma} {$e^{+}e^{-}\rightarrow \pi^+ \pi^- (\gamma)$}
\newcommand {\eegamma} {$e^{+}e^{-}\rightarrow e^+ e^- (\gamma)$}
\newcommand {\taugamma} {$e^{+}e^{-}\rightarrow \tau^+ \tau^- (\gamma)$}
\newcommand {\eemumu} {$e^{+}e^{-}\rightarrow e^{+}e^{-} \mu^+\mu^-$}
\newcommand {\eepipi} {$e^{+}e^{-}\rightarrow e^{+}e^{-} \pi^+\pi^-$}

\newcommand {\fourmu} {$e^{+}e^{-}\rightarrow \mu^{+}\mu^{-} \mu^+\mu^-$}
\newcommand {\foure} {$e^{+}e^{-}\rightarrow e^{+}e^{-} e^+e^-$}
\newcommand {\eett} {$e^{+}e^{-}\rightarrow e^{+}e^{-} \tau^+\tau^-$}

\newcommand {\mmtt} {$e^{+}e^{-}\rightarrow \mu^{+}\mu^{-} \tau^+\tau^-$}

\newcommand {\mumupp} {$e^{+}e^{-}\rightarrow \mu^{+}\mu^{-} \pi^+\pi^-$}

\newcommand {\tautau} {$e^+e^- \to \tau^+\tau^-$}

\newcommand {\zmumu} {$e^{+}e^{-}\rightarrow \mu^+ \mu^- \, Z^\prime$ with $Z^\prime \rightarrow \mu^+\mu^-$}
\newcommand {\Smumu} {$e^{+}e^{-}\rightarrow \mu^+ \mu^- \, S$ with $S \rightarrow \mu^+\mu^-$}
\newcommand {\Xmumu} {$e^{+}e^{-}\rightarrow \mu^+ \mu^- \, X$ with $X \rightarrow \mu^+\mu^-$, $X = Z^\prime, S$}

\newcommand {\lmultau} {$L_{\mu}-L_{\tau}$}

\newcommand {\zprime} {$Z^\prime$}

\newcommand {\zprimemass} {$m_{Z^\prime}$}
\newcommand {\Smass} {$m_{S}$}
\newcommand {\xmass} {$m_{X}$}

\def\invfb   {\ensuremath{\mbox{\,fb}^{-1}}\xspace}

\def\fb   {\ensuremath{\mbox{\,fb}}\xspace}

\newcommand{\gevcc}{\ensuremath{{\mathrm{\,Ge\kern -0.1em V\!/}c^2}}\xspace}
\newcommand{\mevcc}{\ensuremath{{\mathrm{\,Me\kern -0.1em V\!/}c^2}}\xspace}
\newcommand{\gev}{\ensuremath{\mathrm{\,Ge\kern -0.1em V}}\xspace}
\newcommand{\gevc}{\ensuremath{{\mathrm{\,Ge\kern -0.1em V\!/}c}}\xspace}
\newcommand {\mfourmu}{$M({{4\mu}})$}

\newcommand {\Mmumu}{$M({\mu\mu})$}

\def\mumu       {\ensuremath{\mu^+\mu^-}\xspace}

\def\babar{\mbox{\slshape B\kern-0.1em{\smaller A}\kern-0.1em
    B\kern-0.1em{\smaller A\kern-0.2em R}}}
\def\babar{\mbox{\slshape B\kern-0.1em{\smaller A}\kern-0.1em B\kern-0.1em{\smaller A\kern-0.2em R}}}

%% file: supp_content.tex
\renewcommand{\thefigure}{S\arabic{figure}}
\setcounter{figure}{0}

We provide a text file with numerical results of the observed cross sections of \Xmumu,   
as well as of the observed 90\% CL upper limits on the cross section, $g'$, and $g_S$ as functions of the mass.

%% file: main.bbl
\begin{thebibliography}{55}
\expandafter\ifx\csname natexlab\endcsname\relax\def\natexlab#1{#1}\fi
\expandafter\ifx\csname bibnamefont\endcsname\relax
  \def\bibnamefont#1{#1}\fi
\expandafter\ifx\csname bibfnamefont\endcsname\relax
  \def\bibfnamefont#1{#1}\fi
\expandafter\ifx\csname citenamefont\endcsname\relax
  \def\citenamefont#1{#1}\fi
\expandafter\ifx\csname url\endcsname\relax
  \def\url#1{\texttt{#1}}\fi
\expandafter\ifx\csname urlprefix\endcsname\relax\def\urlprefix{URL }\fi
\providecommand{\bibinfo}[2]{#2}
\providecommand{\eprint}[2][]{\url{#2}}

\bibitem[{\citenamefont{Bertone et~al.}(2005)\citenamefont{Bertone, Hooper, and Silk}}]{BERTONE2005279}
\bibinfo{author}{\bibfnamefont{G.}~\bibnamefont{Bertone}}, \bibinfo{author}{\bibfnamefont{D.}~\bibnamefont{Hooper}}, \bibnamefont{and} \bibinfo{author}{\bibfnamefont{J.}~\bibnamefont{Silk}}, \bibinfo{journal}{Phys. Rep.} \textbf{\bibinfo{volume}{405}}, \bibinfo{pages}{279} (\bibinfo{year}{2005}).

\bibitem[{\citenamefont{Bennett et~al.}(2006)}]{PhysRevD.73.072003}
\bibinfo{author}{\bibfnamefont{G.~W.} \bibnamefont{Bennett}} \bibnamefont{et~al.} (\bibinfo{collaboration}{Muon $g\ensuremath{-}2$ Collaboration}), \bibinfo{journal}{Phys. Rev. D} \textbf{\bibinfo{volume}{73}}, \bibinfo{pages}{072003} (\bibinfo{year}{2006}).

\bibitem[{\citenamefont{Aoyama et~al.}(2020)}]{AOYAMA20201}
\bibinfo{author}{\bibfnamefont{T.}~\bibnamefont{Aoyama}} \bibnamefont{et~al.}, \bibinfo{journal}{Phys. Rep.} \textbf{\bibinfo{volume}{887}}, \bibinfo{pages}{1} (\bibinfo{year}{2020}).

\bibitem[{\citenamefont{Aguillard et~al.}(2023)}]{aguillard2023measurement}
\bibinfo{author}{\bibfnamefont{D.~P.} \bibnamefont{Aguillard}} \bibnamefont{et~al.} (\bibinfo{collaboration}{Muon $g\ensuremath{-}2$ Collaboration}), \bibinfo{journal}{Phys. Rev. Lett.} \textbf{\bibinfo{volume}{131}}, \bibinfo{pages}{161802} (\bibinfo{year}{2023}).

\bibitem[{\citenamefont{Borsanyi et~al.}(2021)}]{BMWLattice}
\bibinfo{author}{\bibfnamefont{S.}~\bibnamefont{Borsanyi}} \bibnamefont{et~al.}, Nature(London), \textbf{\bibinfo{volume}{593}} (\bibinfo{year}{2021}).

\bibitem[{\citenamefont{Lees et~al.}(2013)}]{BaBar:2013mob}
\bibinfo{author}{\bibfnamefont{J.~P.} \bibnamefont{Lees}} \bibnamefont{et~al.} (\bibinfo{collaboration}{\textit{BABAR}\ Collaboration}), \bibinfo{journal}{Phys. Rev. D} \textbf{\bibinfo{volume}{88}}, \bibinfo{pages}{072012} (\bibinfo{year}{2013}).

\bibitem[{\citenamefont{Aaij et~al.}(2018)}]{LHCb:2017rln}
\bibinfo{author}{\bibfnamefont{R.}~\bibnamefont{Aaij}} \bibnamefont{et~al.} (\bibinfo{collaboration}{LHCb Collaboration}), \bibinfo{journal}{Phys. Rev. D} \textbf{\bibinfo{volume}{97}}, \bibinfo{pages}{072013} (\bibinfo{year}{2018}).

\bibitem[{\citenamefont{Caria et~al.}(2020)}]{Belle:2019rba}
\bibinfo{author}{\bibfnamefont{G.}~\bibnamefont{Caria}} \bibnamefont{et~al.} (\bibinfo{collaboration}{Belle Collaboration}), \bibinfo{journal}{Phys. Rev. Lett.} \textbf{\bibinfo{volume}{124}}, \bibinfo{pages}{161803} (\bibinfo{year}{2020}).

\bibitem[{\citenamefont{Sala and Straub}(2017)}]{SALA2017205}
\bibinfo{author}{\bibfnamefont{F.}~\bibnamefont{Sala}} \bibnamefont{and} \bibinfo{author}{\bibfnamefont{D.~M.} \bibnamefont{Straub}}, \bibinfo{journal}{Phys. Lett. B} \textbf{\bibinfo{volume}{774}}, \bibinfo{pages}{205} (\bibinfo{year}{2017}).

\bibitem[{\citenamefont{Chen and Nomura}(2018)}]{Chen:2017usq}
\bibinfo{author}{\bibfnamefont{C.-H.} \bibnamefont{Chen}} \bibnamefont{and} \bibinfo{author}{\bibfnamefont{T.}~\bibnamefont{Nomura}}, \bibinfo{journal}{Phys. Lett. B} \textbf{\bibinfo{volume}{777}}, \bibinfo{pages}{420} (\bibinfo{year}{2018}).

\bibitem[{\citenamefont{Greljo et~al.}(2021)\citenamefont{Greljo, Stangl, and {Eller Thomsen}}}]{Greljo:2021xmg}
\bibinfo{author}{\bibfnamefont{A.}~\bibnamefont{Greljo}}, \bibinfo{author}{\bibfnamefont{P.}~\bibnamefont{Stangl}}, \bibnamefont{and} \bibinfo{author}{\bibfnamefont{A.}~\bibnamefont{{Eller Thomsen}}}, \bibinfo{journal}{Phys. Lett. B} \textbf{\bibinfo{volume}{820}}, \bibinfo{pages}{136554} (\bibinfo{year}{2021}).

\bibitem[{\citenamefont{He et~al.}(1991)\citenamefont{He, Joshi, Lew, and Volkas}}]{PhysRevD.43.R22}
\bibinfo{author}{\bibfnamefont{X.~G.} \bibnamefont{He}}, \bibinfo{author}{\bibfnamefont{G.~C.} \bibnamefont{Joshi}}, \bibinfo{author}{\bibfnamefont{H.}~\bibnamefont{Lew}}, \bibnamefont{and} \bibinfo{author}{\bibfnamefont{R.~R.} \bibnamefont{Volkas}}, \bibinfo{journal}{Phys. Rev. D} \textbf{\bibinfo{volume}{43}}, \bibinfo{pages}{R22} (\bibinfo{year}{1991}).

\bibitem[{\citenamefont{Shuve and Yavin}(2014)}]{Shuve:2014doa}
\bibinfo{author}{\bibfnamefont{B.}~\bibnamefont{Shuve}} \bibnamefont{and} \bibinfo{author}{\bibfnamefont{I.}~\bibnamefont{Yavin}}, \bibinfo{journal}{Phys. Rev. D} \textbf{\bibinfo{volume}{89}}, \bibinfo{pages}{113004} (\bibinfo{year}{2014}).

\bibitem[{\citenamefont{Altmannshofer et~al.}(2016)\citenamefont{Altmannshofer, Gori, Profumo, and Queiroz}}]{Altmannshofer:2016jzy}
\bibinfo{author}{\bibfnamefont{W.}~\bibnamefont{Altmannshofer}}, \bibinfo{author}{\bibfnamefont{S.}~\bibnamefont{Gori}}, \bibinfo{author}{\bibfnamefont{S.}~\bibnamefont{Profumo}}, \bibnamefont{and} \bibinfo{author}{\bibfnamefont{F.~S.} \bibnamefont{Queiroz}}, \bibinfo{journal}{J. High Energy Phys.} \textbf{\bibinfo{volume}{12}}, (\bibinfo{year}{2016}) \bibinfo{pages}{106}.

\bibitem[{\citenamefont{Harris et~al.}(2022)\citenamefont{Harris, Schuster, and Zupan}}]{harris2022snowmass}
\bibinfo{author}{\bibfnamefont{P.}~\bibnamefont{Harris}}, \bibinfo{author}{\bibfnamefont{P.}~\bibnamefont{Schuster}}, \bibnamefont{and} \bibinfo{author}{\bibfnamefont{J.}~\bibnamefont{Zupan}}, \eprint{arXiv:2207.08990}.

\bibitem[{\citenamefont{Gori et~al.}(2022)\citenamefont{Gori, Williams, Ilten, Tran, Krnjaic, Toro, Batell, Blinov, Hearty, McGehee et~al.}}]{gori2022dark}
\bibinfo{author}{\bibfnamefont{S.}~\bibnamefont{Gori}}, \bibinfo{author}{\bibfnamefont{M.}~\bibnamefont{Williams}}, \bibinfo{author}{\bibfnamefont{P.}~\bibnamefont{Ilten}}, \bibinfo{author}{\bibfnamefont{N.}~\bibnamefont{Tran}}, \bibinfo{author}{\bibfnamefont{G.}~\bibnamefont{Krnjaic}}, \bibinfo{author}{\bibfnamefont{N.}~\bibnamefont{Toro}}, \bibinfo{author}{\bibfnamefont{B.}~\bibnamefont{Batell}}, \bibinfo{author}{\bibfnamefont{N.}~\bibnamefont{Blinov}}, \bibinfo{author}{\bibfnamefont{C.}~\bibnamefont{Hearty}}, \bibinfo{author}{\bibfnamefont{R.}~\bibnamefont{McGehee}} \bibnamefont{et~al.}, \eprint{arXiv:2209.04671}.

\bibitem[{\citenamefont{Forbes et~al.}(2023)\citenamefont{Forbes, Herwig, Kahn, Krnjaic, Suarez, Tran, and Whitbeck}}]{forbes2023new}
\bibinfo{author}{\bibfnamefont{D.}~\bibnamefont{Forbes}}, \bibinfo{author}{\bibfnamefont{C.}~\bibnamefont{Herwig}}, \bibinfo{author}{\bibfnamefont{Y.}~\bibnamefont{Kahn}}, \bibinfo{author}{\bibfnamefont{G.}~\bibnamefont{Krnjaic}}, \bibinfo{author}{\bibfnamefont{C.~M.} \bibnamefont{Suarez}}, \bibinfo{author}{\bibfnamefont{N.}~\bibnamefont{Tran}}, \bibnamefont{and} \bibinfo{author}{\bibfnamefont{A.}~\bibnamefont{Whitbeck}}, \bibinfo{journal}{Phys. Rev. D} \textbf{\bibinfo{volume}{107}}, \bibinfo{pages}{116026} (\bibinfo{year}{2023}).

\bibitem[{\citenamefont{Capdevilla et~al.}(2022)\citenamefont{Capdevilla, Curtin, Kahn, and Krnjaic}}]{Capdevilla_2022}
\bibinfo{author}{\bibfnamefont{R.}~\bibnamefont{Capdevilla}}, \bibinfo{author}{\bibfnamefont{D.}~\bibnamefont{Curtin}}, \bibinfo{author}{\bibfnamefont{Y.}~\bibnamefont{Kahn}}, \bibnamefont{and} \bibinfo{author}{\bibfnamefont{G.}~\bibnamefont{Krnjaic}}, \bibinfo{journal}{J. High Energy Phys.} \textbf{\bibinfo{volume}{04}}, (\bibinfo{year}{2022}) \bibinfo{pages}{129}.

\bibitem[{\citenamefont{Lees et~al.}(2016)}]{TheBABAR:2016rlg}
\bibinfo{author}{\bibfnamefont{J.~P.} \bibnamefont{Lees}} \bibnamefont{et~al.} (\bibinfo{collaboration}{\textit{BABAR}\ Collaboration}), \bibinfo{journal}{Phys. Rev. D} \textbf{\bibinfo{volume}{94}}, \bibinfo{pages}{011102} (\bibinfo{year}{2016}).

\bibitem[{\citenamefont{Czank et~al.}(2022)}]{PhysRevD.106.012003}
\bibinfo{author}{\bibfnamefont{T.}~\bibnamefont{Czank}} \bibnamefont{et~al.} (\bibinfo{collaboration}{Belle Collaboration}), \bibinfo{journal}{Phys. Rev. D} \textbf{\bibinfo{volume}{106}}, \bibinfo{pages}{012003} (\bibinfo{year}{2022}).

\bibitem[{\citenamefont{Sirunyan et~al.}(2019)}]{cms}
\bibinfo{author}{\bibfnamefont{A.}~\bibnamefont{Sirunyan}} \bibnamefont{et~al.} (\bibinfo{collaboration}{CMS Collaboration}), \bibinfo{journal}{Phys. Lett. B} \textbf{\bibinfo{volume}{792}}, \bibinfo{pages}{345} (\bibinfo{year}{2019}).

\bibitem[{\citenamefont{Adachi et~al.}(2020)}]{PhysRevLett.124.141801}
\bibinfo{author}{\bibfnamefont{I.}~\bibnamefont{Adachi}} \bibnamefont{et~al.} (\bibinfo{collaboration}{Belle II Collaboration}), \bibinfo{journal}{Phys. Rev. Lett.} \textbf{\bibinfo{volume}{124}}, \bibinfo{pages}{141801} (\bibinfo{year}{2020}).

\bibitem[{\citenamefont{Adachi et~al.}(2023{\natexlab{a}})}]{zpinvis_update}
\bibinfo{author}{\bibfnamefont{I.}~\bibnamefont{Adachi}} \bibnamefont{et~al.} (\bibinfo{collaboration}{Belle II Collaboration}), \bibinfo{journal}{Phys. Rev. Lett.} \textbf{\bibinfo{volume}{130}}, \bibinfo{pages}{231801} (\bibinfo{year}{2023}{\natexlab{a}}).

\bibitem[{\citenamefont{Andreev et~al.}(2022)}]{PhysRevD.106.032015}
\bibinfo{author}{\bibfnamefont{Y.~M.} \bibnamefont{Andreev}} \bibnamefont{et~al.} (\bibinfo{collaboration}{NA64 Collaboration}), \bibinfo{journal}{Phys. Rev. D} \textbf{\bibinfo{volume}{106}}, \bibinfo{pages}{032015} (\bibinfo{year}{2022}).

\bibitem[{\citenamefont{Adachi et~al.}(2023{\natexlab{b}})}]{xtotautau}
\bibinfo{author}{\bibnamefont{Adachi}} \bibnamefont{et~al.} (\bibinfo{collaboration}{Belle II Collaboration}), \bibinfo{journal}{Phys. Rev. Lett.} \textbf{\bibinfo{volume}{131}}, \bibinfo{pages}{121802} (\bibinfo{year}{2023}{\natexlab{b}}).

\bibitem[{\citenamefont{Abe et~al.}(2010)}]{Abe:2010sj}
\bibinfo{author}{\bibfnamefont{T.}~\bibnamefont{Abe}} \bibnamefont{et~al.}, \eprint{arXiv:1011.0352}.

\bibitem[{\citenamefont{Kou et~al.}(2019)}]{ref:b2tip}
\bibinfo{author}{\bibfnamefont{E.}~\bibnamefont{Kou}} \bibnamefont{et~al.}, \bibinfo{journal}{Prog. Theor. Exp. Phys.} \textbf{\bibinfo{volume}{2019}}, \bibinfo{pages}{123C01} (\bibinfo{year}{2019}); \bibinfo{note}{029201(E) (2020)}
%\bibinfo{note}{{E}rratum: \url{https://doi.org/10.1093/ptep/ptaa008}}.

\bibitem[{\citenamefont{Akai et~al.}(2018)\citenamefont{Akai, Furukawa, and Koiso}}]{superkekb}
\bibinfo{author}{\bibfnamefont{K.}~\bibnamefont{Akai}}, \bibinfo{author}{\bibfnamefont{K.}~\bibnamefont{Furukawa}}, \bibnamefont{and} \bibinfo{author}{\bibfnamefont{H.}~\bibnamefont{Koiso}} (\bibinfo{collaboration}{SuperKEKB Accelerator Team}), \bibinfo{journal}{Nucl. Instrum. Methods Phys. Res., Sect. A} \textbf{\bibinfo{volume}{907}}, \bibinfo{pages}{188} (\bibinfo{year}{2018}).

\bibitem[{\citenamefont{Abudin{\'{e}}n et~al.}(2020)}]{lumi}
\bibinfo{author}{\bibfnamefont{F.}~\bibnamefont{Abudin{\'{e}}n}} \bibnamefont{et~al.} (\bibinfo{collaboration}{Belle II Collaboration}), \bibinfo{journal}{Chin. Phys. C} \textbf{\bibinfo{volume}{44}}, \bibinfo{pages}{021001} (\bibinfo{year}{2020}).

\bibitem[{\citenamefont{Alwall et~al.}(2014)}]{Alwall2014}
\bibinfo{author}{\bibfnamefont{J.}~\bibnamefont{Alwall}} \bibnamefont{et~al.}, \bibinfo{journal}{J. High Energy Phys.} \textbf{\bibinfo{volume}{07}}, (\bibinfo{year}{2014}) \bibinfo{pages}{079}.

\bibitem[{\citenamefont{Li and Yan}(2018)}]{isr-plugin}
\bibinfo{author}{\bibfnamefont{Q.}~\bibnamefont{Li}} \bibnamefont{and} \bibinfo{author}{\bibfnamefont{Q.-S.} \bibnamefont{Yan}}, \eprint{arXiv:1804.00125}.

\bibitem[{\citenamefont{Berends et~al.}(1985)\citenamefont{Berends, Daverveldt, and Kleiss}}]{ref:fourlepton}
\bibinfo{author}{\bibfnamefont{F.}~\bibnamefont{Berends}}, \bibinfo{author}{\bibfnamefont{P.}~\bibnamefont{Daverveldt}}, \bibnamefont{and} \bibinfo{author}{\bibfnamefont{R.}~\bibnamefont{Kleiss}}, \bibinfo{journal}{Nucl. Phys. B253}, \bibinfo{pages}{441} (\bibinfo{year}{1985}).
%\textbf{\bibinfo{volume}{253}}

\bibitem[{\citenamefont{Jadach et~al.}(2000)\citenamefont{Jadach, Ward, and W\k{a}s}}]{ref:kkmc}
\bibinfo{author}{\bibfnamefont{S.}~\bibnamefont{Jadach}}, \bibinfo{author}{\bibfnamefont{B.~F.~L.} \bibnamefont{Ward}}, \bibnamefont{and} \bibinfo{author}{\bibfnamefont{Z.}~\bibnamefont{W\k{a}s}}, \bibinfo{journal}{Comput. Phys. Commun.} \textbf{\bibinfo{volume}{130}}, \bibinfo{pages}{260} (\bibinfo{year}{2000}).

\bibitem[{\citenamefont{Davidson et~al.}(2012)\citenamefont{Davidson, Nanava, Przedzinski, Richter-W\k{a}s, and W\k{a}s}}]{ref:tauola}
\bibinfo{author}{\bibfnamefont{N.}~\bibnamefont{Davidson}}, \bibinfo{author}{\bibfnamefont{G.}~\bibnamefont{Nanava}}, \bibinfo{author}{\bibfnamefont{T.}~\bibnamefont{Przedzinski}}, \bibinfo{author}{\bibfnamefont{E.}~\bibnamefont{Richter-W\k{a}s}}, \bibnamefont{and} \bibinfo{author}{\bibfnamefont{Z.}~\bibnamefont{W\k{a}s}}, \bibinfo{journal}{Comput. Phys. Commun.} \textbf{\bibinfo{volume}{183}}, \bibinfo{pages}{821} (\bibinfo{year}{2012}).

\bibitem[{\citenamefont{Uehara}(2013)}]{uehara2013treps}
\bibinfo{author}{\bibfnamefont{S.}~\bibnamefont{Uehara}}, \eprint{arXiv:1310.0157}.

\bibitem[{\citenamefont{Czyż et~al.}(2013)\citenamefont{Czyż, Gunia, and Kühn}}]{ref:phokhara}
\bibinfo{author}{\bibfnamefont{H.}~\bibnamefont{Czyż}}, \bibinfo{author}{\bibfnamefont{M.}~\bibnamefont{Gunia}}, \bibnamefont{and} \bibinfo{author}{\bibfnamefont{J.~H.} \bibnamefont{Kühn}}, \bibinfo{journal}{J. High Energy Phys.} \textbf{\bibinfo{volume}{08}}, (\bibinfo{year}{2013}) \bibinfo{pages}{110}.

\bibitem[{\citenamefont{Balossini et~al.}(2008)\citenamefont{Balossini, Bignamini, Calame, Montagna, Nicrosini, and Piccinini}}]{ref:babayaga}
\bibinfo{author}{\bibfnamefont{G.}~\bibnamefont{Balossini}}, \bibinfo{author}{\bibfnamefont{C.}~\bibnamefont{Bignamini}}, \bibinfo{author}{\bibfnamefont{C.~M.~C.} \bibnamefont{Calame}}, \bibinfo{author}{\bibfnamefont{G.}~\bibnamefont{Montagna}}, \bibinfo{author}{\bibfnamefont{O.}~\bibnamefont{Nicrosini}}, \bibnamefont{and} \bibinfo{author}{\bibfnamefont{F.}~\bibnamefont{Piccinini}}, \bibinfo{journal}{Phys. Lett. B} \textbf{\bibinfo{volume}{663}}, \bibinfo{pages}{209} (\bibinfo{year}{2008}).

\bibitem[{\citenamefont{Sjöstrand et~al.}(2015)}]{pythia8}
\bibinfo{author}{\bibfnamefont{T.}~\bibnamefont{Sjöstrand}} \bibnamefont{et~al.}, \bibinfo{journal}{Comput. Phys. Commun.} \textbf{\bibinfo{volume}{191}}, \bibinfo{pages}{159} (\bibinfo{year}{2015}).

\bibitem[{\citenamefont{Lange}(2001)}]{evtgen}
\bibinfo{author}{\bibfnamefont{D.~J.} \bibnamefont{Lange}}, \bibinfo{journal}{Nucl. Instrum. Methods Phys. Res. Sect. A} \textbf{\bibinfo{volume}{462}}, \bibinfo{pages}{152} (\bibinfo{year}{2001}).

\bibitem[{\citenamefont{Barberio et~al.}(1991)\citenamefont{Barberio, van Eijk, and W\k{a}s}}]{Barberio:1990ms}
\bibinfo{author}{\bibfnamefont{E.}~\bibnamefont{Barberio}}, \bibinfo{author}{\bibfnamefont{B.}~\bibnamefont{van Eijk}}, \bibnamefont{and} \bibinfo{author}{\bibfnamefont{Z.}~\bibnamefont{W\k{a}s}}, \bibinfo{journal}{Comput. Phys. Commun.} \textbf{\bibinfo{volume}{66}}, \bibinfo{pages}{115} (\bibinfo{year}{1991}).

\bibitem[{\citenamefont{Barberio and W\k{a}s}(1994)}]{Barberio:1993qi}
\bibinfo{author}{\bibfnamefont{E.}~\bibnamefont{Barberio}} \bibnamefont{and} \bibinfo{author}{\bibfnamefont{Z.}~\bibnamefont{W\k{a}s}}, \bibinfo{journal}{Comput. Phys. Commun.} \textbf{\bibinfo{volume}{79}}, \bibinfo{pages}{291} (\bibinfo{year}{1994}).

\bibitem[{\citenamefont{Agostinelli et~al.}(2003)}]{ref:geant4}
\bibinfo{author}{\bibfnamefont{S.}~\bibnamefont{Agostinelli}} \bibnamefont{et~al.} (\bibinfo{collaboration}{\textsc{Geant4}}), \bibinfo{journal}{Nucl. Instrum. Methods Phys. Res. Sect A} \textbf{\bibinfo{volume}{506}}, \bibinfo{pages}{250} (\bibinfo{year}{2003}).

\bibitem[{\citenamefont{Kuhr et~al.}(2019)\citenamefont{Kuhr, Pulvermacher, Ritter, Hauth, and Braun}}]{basf2}
\bibinfo{author}{\bibfnamefont{T.}~\bibnamefont{Kuhr}}, \bibinfo{author}{\bibfnamefont{C.}~\bibnamefont{Pulvermacher}}, \bibinfo{author}{\bibfnamefont{M.}~\bibnamefont{Ritter}}, \bibinfo{author}{\bibfnamefont{T.}~\bibnamefont{Hauth}}, \bibnamefont{and} \bibinfo{author}{\bibfnamefont{N.}~\bibnamefont{Braun}} (\bibinfo{collaboration}{Belle II Framework Software Group}), \bibinfo{journal}{Comput. Software Big Sci.} \textbf{\bibinfo{volume}{3}}, \bibinfo{pages}{1} (\bibinfo{year}{2019}).

\bibitem[{bas()}]{basf2-zenodo}
\bibinfo{title}{Belle {I}{I} {A}nalysis {S}oftware {F}ramework (basf2)}, \bibinfo{howpublished}{\url{https://doi.org/10.5281/zenodo.5574115}}.

\bibitem[{\citenamefont{Bertacchi et~al.}(2021)}]{tracking}
\bibinfo{author}{\bibfnamefont{V.}~\bibnamefont{Bertacchi}} \bibnamefont{et~al.} (\bibinfo{collaboration}{Belle II Tracking Group}), \bibinfo{journal}{Comput. Phys. Commun.} \textbf{\bibinfo{volume}{259}}, \bibinfo{pages}{107610} (\bibinfo{year}{2021}).

\bibitem[{\citenamefont{A.~Hoecker et~al.}(2009)}]{hoecker2009tmva}
\bibinfo{author}{\bibfnamefont{A.}~\bibnamefont{A.~Hoecker}} \bibnamefont{et~al.}, \eprint{arXiv:physics/0703039}.

\bibitem[{\citenamefont{Punzi}(2003)}]{Punzi:2003bu}
\bibinfo{author}{\bibfnamefont{G.}~\bibnamefont{Punzi}}, \bibinfo{journal}{eConf} \textbf{\bibinfo{volume}{C030908}}, \bibinfo{pages}{MODT002}, \eprint{arXiv:physics/0308063}.

\bibitem[{\citenamefont{Skwarnicki}(1986)}]{Skwarnicki:1986xj}
\bibinfo{author}{\bibfnamefont{T.}~\bibnamefont{Skwarnicki}}, Ph.D. thesis, \bibinfo{school}{Cracow, INP}, \bibinfo{year}{1986}.

\bibitem[{\citenamefont{Efron and Tibshirani}(1994)}]{efron1994introduction}
\bibinfo{author}{\bibfnamefont{B.}~\bibnamefont{Efron}} \bibnamefont{and} \bibinfo{author}{\bibfnamefont{R.~J.} \bibnamefont{Tibshirani}}, \emph{\bibinfo{title}{An Introduction to the Bootstrap}} (\bibinfo{publisher}{Chapman and Hall/CRC}, \bibinfo{address}{New York}, \bibinfo{year}{1994}).

\bibitem[{\citenamefont{Cowan et~al.}(2011)\citenamefont{Cowan, Cranmer, Gross, and Vitells}}]{Cowan:2010js}
\bibinfo{author}{\bibfnamefont{G.}~\bibnamefont{Cowan}}, \bibinfo{author}{\bibfnamefont{K.}~\bibnamefont{Cranmer}}, \bibinfo{author}{\bibfnamefont{E.}~\bibnamefont{Gross}}, \bibnamefont{and} \bibinfo{author}{\bibfnamefont{O.}~\bibnamefont{Vitells}}, \bibinfo{journal}{Eur. Phys. J. C} \textbf{\bibinfo{volume}{71}}, \bibinfo{pages}{1554} (\bibinfo{year}{2011}); \bibinfo{note}{2501(E), (2013)}.

\bibitem[{\citenamefont{Gross and Vitells}(2010)}]{Gross:2010qma}
\bibinfo{author}{\bibfnamefont{E.}~\bibnamefont{Gross}} \bibnamefont{and} \bibinfo{author}{\bibfnamefont{O.}~\bibnamefont{Vitells}}, \bibinfo{journal}{Eur. Phys. J. C} \textbf{\bibinfo{volume}{70}}, \bibinfo{pages}{525} (\bibinfo{year}{2010}).

\bibitem[{\citenamefont{Read}(2002)}]{CLS}
\bibinfo{author}{\bibfnamefont{A.~L.} \bibnamefont{Read}}, \bibinfo{journal}{J. Phys. G} \textbf{\bibinfo{volume}{28}}, \bibinfo{pages}{2693} (\bibinfo{year}{2002}).

\bibitem[{\citenamefont{Altmannshofer et~al.}(2014)\citenamefont{Altmannshofer, Gori, Pospelov, and Yavin}}]{PhysRevLett.113.091801}
\bibinfo{author}{\bibfnamefont{W.}~\bibnamefont{Altmannshofer}}, \bibinfo{author}{\bibfnamefont{S.}~\bibnamefont{Gori}}, \bibinfo{author}{\bibfnamefont{M.}~\bibnamefont{Pospelov}}, \bibnamefont{and} \bibinfo{author}{\bibfnamefont{I.}~\bibnamefont{Yavin}}, \bibinfo{journal}{Phys. Rev. Lett.} \textbf{\bibinfo{volume}{113}}, \bibinfo{pages}{091801} (\bibinfo{year}{2014}).

\bibitem[{\citenamefont{Bellini et~al.}(2011)}]{PhysRevLett.107.141302}
\bibinfo{author}{\bibfnamefont{G.}~\bibnamefont{Bellini}} \bibnamefont{et~al.} (\bibinfo{collaboration}{Borexino Collaboration}), \bibinfo{journal}{Phys. Rev. Lett.} \textbf{\bibinfo{volume}{107}}, \bibinfo{pages}{141302} (\bibinfo{year}{2011}).

\bibitem[{\citenamefont{Kamada et~al.}(2018)\citenamefont{Kamada, Kaneta, Yanagi, and Yu}}]{kamada2018self}
\bibinfo{author}{\bibfnamefont{A.}~\bibnamefont{Kamada}}, \bibinfo{author}{\bibfnamefont{K.}~\bibnamefont{Kaneta}}, \bibinfo{author}{\bibfnamefont{K.}~\bibnamefont{Yanagi}}, \bibnamefont{and} \bibinfo{author}{\bibfnamefont{H.-B.} \bibnamefont{Yu}}, \bibinfo{journal}{J. High Energy Phys.} \textbf{\bibinfo{volume}{2018}}, (\bibinfo{year}{2018}) \bibinfo{pages}{117}.

\bibitem{supp} See Supplemental Material at \url{https://journals.aps.org/prd/abstract/10.1103/PhysRevD.109.112015#supplemental} for numerical results of the observed cross sections of $X\to\mu\mu$, as well as of the observed 90\% CL upper limits on the cross section, $g^{\prime}$, and $g_S$ as functions of the mass.

\end{thebibliography}
